\documentclass[floats,floatfix,showpacs,amssymb,prd,twocolumn,superscriptaddress,nofootinbib,nolongbibliography,reprint,aps]{revtex4-2}

\usepackage{silence}
\usepackage{graphicx,epsf,epsfig,amssymb,physics,microtype}
\usepackage{bm}
\usepackage{color}
\usepackage{amsfonts,amsmath,mathrsfs,textcomp,gensymb}
\usepackage[breaklinks,bookmarksopen=true]{hyperref}
\usepackage[capitalise]{cleveref}

\usepackage{natbib}
\usepackage{prd_macros}
\usepackage{multirow}
\usepackage{rotating,array}
\usepackage[normalem]{ulem}
\usepackage{dcolumn}
\usepackage{braket}
\usepackage{appendix}
\usepackage{caption}
\usepackage{subcaption}
\usepackage{comment}
\usepackage[dvipsnames]{xcolor}
\definecolor{linkcolor}{rgb}{0.0,0.3,0.5}
\usepackage[all]{hypcap}
\usepackage[T1]{fontenc}
\usepackage[utf8]{inputenc}
\usepackage{tabularx}

\graphicspath{{figures/}}
\usepackage [english]{babel}
\usepackage [autostyle, english = american]{csquotes}
\MakeOuterQuote{"}

\newcommand{\mk}[1] {\textcolor{orange}{#1 \textsc{/mk/}}}
\newcommand{\ts}[1] {\textcolor{red}{#1 \textsc{/ts/}}}

\newcommand{\done}[1]{}

\newcommand{\dallas}{\affiliation{Department of Physics, The University of Texas at Dallas, Richardson, Texas 75080, USA}}

\newcommand\orcid[1]{\href{https://orcid.org/#1}{$\!$\includegraphics[scale=0.006]{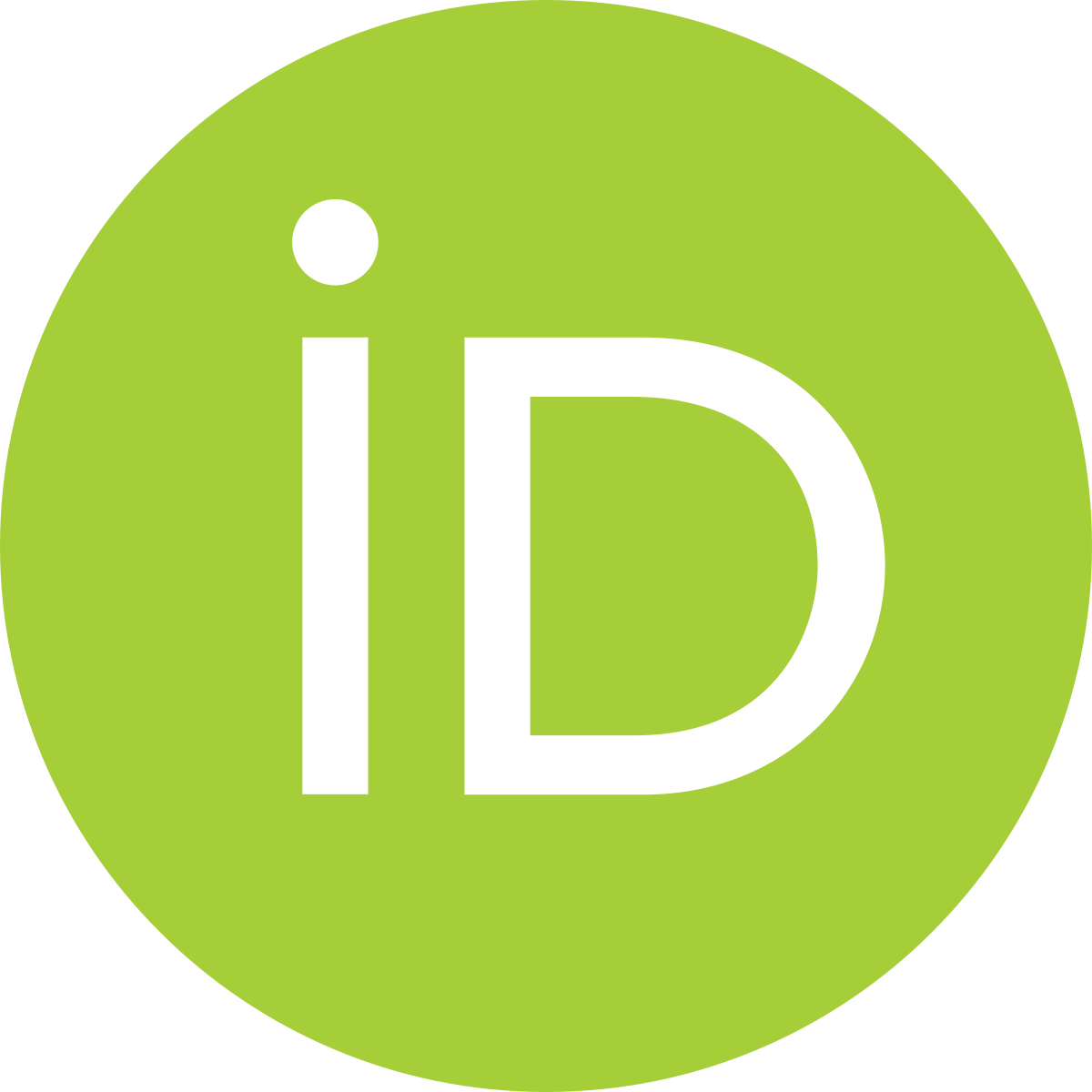} $\!\!$}}

\newcommand{\tdel}{\Delta t_d}
\newcommand{\hL}{\tilde{h}_L}
\newcommand{\hUL}{\tilde{h}_{UL}}

\newcommand{\hS}{\tilde{h}_s}   
\newcommand{\hT}{\tilde{h}_t}   
\newcommand{\BL}{B_L}
\newcommand{\BNP}{B_{NP}}
\newcommand{\BRP}{B_{RP}}
\newcommand{\BS}{B_s}
\newcommand{\BT}{B_t}
\newcommand{\PhiL}{\Phi_L}

\newcommand{\PhiNP}{\Phi_{NP}}
\newcommand{\PhiRP}{\Phi_{RP}}

\newcommand{\epsST}{\epsilon(\hS, \hT)}
\newcommand{\epsP}{\epsilon_P}
\newcommand{\epsNP}{\epsilon_{NP}}
\newcommand{\epsRP}{\epsilon_{RP}}

\newcommand{\gammaP}{\gamma_P}
\newcommand{\OmLJ}{\Omega_{LJ}}
\newcommand{\PhiLJ}{\Phi_{LJ}}
\newcommand{\thLJ}{\theta_{LJ}}
\newcommand{\iotaJN}{\iota_{JN}}
\newcommand{\Omtil}{\tilde{\Omega}}
\newcommand{\thtil}{\tilde{\theta}}
\newcommand{\Omtilbest}{\tilde{\Omega}_{\mathrm{best}}}
\newcommand{\thtilbest}{\tilde{\theta}_{\mathrm{best}}}
\newcommand{\fcut}{f_{\mathrm{cut}}}
\newcommand{\fmin}{f_{\mathrm{min}}}
\newcommand{\Mcs}{\mathcal{M}_s}

\newcommand{\Nfringe}{N_{\mathrm{fringe}}}

\begin{document}

\title{Distinguishing lensing and precessional modulation in binary black-hole inspiral waveforms}

\author{Tien N. Nguyen-Vo}
\thanks{Co-lead}
\email{tvn6@cornell.edu} 
\dallas
\affiliation{Department of Physics, Cornell University, Ithaca, New York 14853, USA}

\author{Tamanjyot Singh \orcid{0009-0006-4040-4407}}
\thanks{Co-lead}
\email{ftamanj1@jh.edu} 
\dallas
\affiliation{William H. Miller III Department of Physics and Astronomy, Johns Hopkins University, \\ 3400 North Charles Street, Baltimore, Maryland, 21218, USA}

\author{Benjamin McKallip}
\email{mckallipb@southwestern.edu}
\dallas
\affiliation{Department of Physics, Southwestern University, Georgetown, Texas 78626, USA}

\author{Michael Kesden \orcid{0000-0002-5987-1471}}
\email{kesden@utdallas.edu}
\dallas

\author{Lindsay King \orcid{0000-0001-5732-3538}}
\email{Lindsay.King@utdallas.edu}
\dallas

\date{\today}

\begin{abstract}
Binary black holes (BBHs) emit gravitational waves (GWs) as they inspiral towards merger.
These GWs can be gravitationally lensed by large-scale structure along the line of sight, potentially creating multiple images of the same source with fixed time delays determined by the lensing geometry.
As the BBHs inspiral, the GW frequency increases, leading to successive constructive and destructive interference between the multiple images.
BBHs also have spins $\mathbf{S}_i$ that may be misaligned with their orbital angular momentum $\mathbf{L}$.
As the BBHs inspiral, these misaligned spins cause $\mathbf{L}$ to precess about the total angular momentum $\mathbf{J}$, modulating the GW emission similar to pulsar emission resulting from a misaligned jet rotating in and out of the line of sight.
We investigate the ability of a single L-shaped GW detector to distinguish between these two sources of modulation.
We find that precessional modulation can mimic the lensing modulation between two images with comparable magnifications when the time delay between the images is short enough that fewer than three interference fringes occur during the time the GW signal spends in the sensitivity band of the detector.
As strong lensing is rare for GW sources at moderate redshift while misaligned spins are common for BBHs produced in certain formation channels, ruling out precessional modulation is essential to identifying genuinely lensed systems.
\end{abstract}

\maketitle

\section{Introduction} \label{sec: Intro}
Electromagnetic (EM) radiation propagates along null geodesics in curved spacetime.
The resulting deflection of light and other EM radiation by intervening massive objects such as galaxies, known as gravitational lensing, is a well-established phenomenon.
Einstein first derived the equations for gravitational lensing and multiple imaging by a point mass in 1912 within the framework of general relativity~\cite{einstein1995zurich}, later publishing these results in 1936~\cite{Einstein1936}.
An earlier published calculation of lensing by a point mass was presented by Khvolson in 1924~\cite{chwolson1924doppelsterne}.
In 1979, the discovery of the first strongly lensed source, the double quasar Q0957+561, confirmed that multiple, magnified images can be produced by the strong gravitational lensing of a single background source~\cite{1979Natur.279..381W}.
Since then, hundreds of strongly lensed systems have been identified across the EM spectrum~\cite{2003MNRAS.341...13B, 2008ApJ...682..964B}.
Gravitational lensing has become a standard tool in astrophysics and cosmology, enabling precision studies of dark matter, distant sources, and cosmological parameters~\cite{Wambs_rev_1998LRR.....1...12W, 2024SSRv..220...58V, Cluster_rev_2024SSRv..220...19N}.

Gravitational waves (GWs) also propagate along null geodesics and are susceptible to the same deflections due to spacetime curvature around massive objects~\cite{Lawrence1971,Nakamura1998,Nakamura1999,TakahashiNakamura2003,Oguri2018}.
Analogous to strong EM lensing, multiple, magnified, time-delayed images of a GW source signal can be created.
However, unlike most EM sources, binary black holes (BBHs) emit GWs coherently, leading to interference between multiple images of the same source, provided that the duration of the source is longer than the time delay between images~\cite{Ali2023}.
As BBHs inspiral, the frequencies of the GWs they emit increase, leading to increasing phase shifts between multiple images with fixed time delays.
These changing phase shifts lead to alternating constructive and destructive interference, modulating the amplitude and phase of the GW strain observed by ground- and space-based detectors.

After more than a decade of GW observations, and almost 400 confirmed GW sources in the LIGO-Virgo-KAGRA (LVK) data~\cite{GWTC5}, several candidate strongly lensed sources have been proposed, but none have been confirmed~\cite{2025arXiv251216347T}.
Most of the searches for multiply imaged GW sources have been based on the framework of 
\citeauthor{2018arXiv180707062H}
~\cite{2018arXiv180707062H}, and have focused on the regime where the source duration is much shorter than the time delay between images.
In this regime, the expected signature of strong lensing is distinct repeated images with time delays on the order of days to months, produced by galaxies and groups of galaxies.
Several candidate lensed GW pairs have been proposed, but no confirmed lensed GW detection exists yet in LVK data~\cite{Abbott2024_LensingO3,Janquart2023_LensingFollowup,Janquart2021,Janquart2023,Li2023,Lo2023,Bianconi2023,Chakraborty2026}.
The most promising candidate to date is the massive GW source GW231123~\cite{2025ApJ_GW23}.
\citeauthor{2025_Goyal_GW231123}~\cite{2025_Goyal_GW231123} suggested that diffraction by a point-mass lens of several hundred solar masses, with or without a galaxy-mass macrolens, better reproduces the data than an unlensed waveform. However, employing a deep-learning algorithm, less support for the lensing hypothesis was found by~\citeauthor{2025_Chan_GW231123}~\cite{2025_Chan_GW231123}, leaving the status of the candidate unresolved.

Lensing of GW sources has great potential as a new probe of gravitation, astrophysics, and cosmology~\cite{Takahashi2003,Dai2017_Population,Liao2017precision,Wierda2021,Meena2020,Caliskan2024,ChenLu2026}.
Of particular relevance here is that GW lensing may be sensitive to compact objects and low-mass halos in the mass range $\sim 10^2$--$10^{8}\,M_\odot$, bridging the gap between stellar-mass microlensing and galaxy-scale lensing, which is extremely difficult to study directly with EM lensing.

BBH spin precession also modulates GW emission compared to a non-precessing (NP) source.
GW emission is beamed in the direction of the orbital angular momentum $\mathbf{L}$.
If BBH spins are misaligned with the orbital angular momentum, $\mathbf{L}$ will precess about the total angular momentum $\mathbf{J}$ whose direction is generally conserved during the inspiral~\cite{Apostolatos1994,GangardtSteinle2021,Zhao2017}.
This will modulate the amplitude and phase of the observed GW strain as $\mathbf{L}$ precesses in and out of the line of sight~\cite{Apostolatos1994,TamanRP2025}, similar to the pulses observed from a pulsar, a rotating neutron star whose jet is misaligned with its rotational axis.

The goal of this paper is to investigate possible degeneracy between these two sources of GW modulation: gravitational lensing and precession.
This lensing-precession degeneracy has already received some attention~\cite{LiuKim2024,Shan2026}, but our study differs from these in that it employs a new model of regularly precessing (RP) waveforms~\cite{TamanRP2025} parameterized by the dimensionless precession frequency $\Omtil$ and amplitude $\thtil$~\cite{GangardtSteinle2021} rather than the components of individual BBH spins.
As these precession parameters are more transparently connected to the GW modulation than individual spin components, we hypothesize that the lensing-precession degeneracy will also be more transparent in terms of these parameters.

In the simplest case of a two-image lensed source in the geometrical-optics regime~\cite{TakahashiNakamura2003}, the lensed waveform can be fully characterized by the time delay $\tdel$ and flux ratio $I$ between the two images~\cite{Ali2023}.
In the frequency domain, increasing $\tdel$ reduces the separation between interference fringes, while increasing $I$ increases the amplitude of these fringes.
As increasing $\Omtil$ and $\thtil$ have qualitatively similar effects on the GW strain for precessing BBHs, we hypothesize that the lensing-precession degeneracy can most readily be interpreted in terms of these parameters.
We investigate the validity of this hypothesis in the remainder of this paper, finding that it does indeed hold for gravitational waveforms with between roughly one and three interference fringes within the sensitivity band of the GW detector.

In \cref{sec: Methodology}, we present our choice of unlensed, non-precessing waveforms, review how gravitational lensing and precession modulate these waveforms, and describe how we use the mismatch between waveforms to identify degeneracies.
In \cref{sec: Results}, we calculate the minimum mismatch between lensed source waveforms and RP templates as a function of the chirp mass $\Mcs$, time delay $\tdel$, and flux ratio $I$ of the source.
We also explore how the dimensionless precession frequency $\Omtil$ and amplitude $\thtil$ of the best-fitting RP templates vary with these lensing parameters.
A brief summary of our results and their implications is provided in \cref{sec: Discussion}.
In \cref{appendix sec: analytical mismatch}, we examine how lensing-induced interference fringes propagate into fringes in the mismatch between lensed and unlensed NP waveforms as a function of chirp mass $\Mcs$ and time delay $\tdel$.
We also show that minimizing the mismatch with respect to the chirp mass of the template has a negligible effect on our analysis.
In \cref{appendix sec: secular phase}, we investigate how secular phase accumulation in precessing waveforms depends on precession parameters.
Throughout this paper, we use relativists' units in which Newton's gravitational constant and the speed of light equal unity ($G = c = 1$).  

\section{Methodology} 
\label{sec: Methodology}

We begin with a review of the inspiral waveform in the post-Newtonian (PN), quadrupole-moment approximation.

\subsection{Gravitational waveform}
\label{subsec: GW background_LVP}

The frequency-domain strain $\tilde{h}$ received at a ground-based interferometer can be written as \cite{CutlerFlanagan1994, Apostolatos1994, TamanRP2025}:
\begin{equation} \label{eq: hf A and phase_LVP}
\tilde{h}(f) = B e^{\mathit{i}[\Psi(f) - \phi_p - 2\delta\Phi]} \,,
\end{equation}
where $B$ is the GW amplitude given by
\begin{align} 
B &= ACf^{-7/6} \{ 4(\hat{\Vec{L}} \cdot \hat{\Vec{N}})^2 \sin^2(2\psi + \alpha) \notag \\
&\qquad + [1+(\hat{\Vec{L}}\cdot\hat{\Vec{N}})^2 ]^2 \cos^2(2\psi + \alpha) \}^{1/2} \,. \label{eq: hf amplitude_LVP}
\end{align}
We specify the sky location $\hat{\Vec{N}}$ of the GW source by its polar angles $\theta_S, \Phi_S$ with respect to the detector frame $\{ \hat{\Vec{X}}_D, \hat{\Vec{Y}}_D, \hat{\Vec{Z}}_D \}$ in which $\hat{\Vec{X}}_D$ and $\hat{\Vec{Y}}_D$ point along the arms of the L-shaped GW detector.
$\hat{\Vec{L}}$ is the unit vector in the direction of the BBH orbital angular momentum.
$C$ and $\alpha$ are the detector beam-pattern amplitude and phase given as functions of the sky location by \cite{TamanRP2025}
\begin{subequations} \label{eq: BeamPatt_amp_phase_LVP}
\begin{align}
C &= \left[ \frac{1}{4} (1+\cos^2\theta_S)^2 \cos^2 2\Phi_S + \cos^2\theta_S \sin^2 2\Phi_S \right]^{1/2}, \label{eq: BeamPattamp_LVP} \\
\alpha &= \tan^{-1}\left( \frac{2\cos\theta_S \tan 2\Phi_S}{1+\cos^2\theta_S} \right) \,. \label{eq: BeamPattphase_LVP}
\end{align}
\end{subequations}
The polarization angle $\psi$ between the principal $+$ direction and the direction of constant azimuth is given by \cite{TamanRP2025, Apostolatos1994}
\begin{equation} \label{eq: polarization angle psi_LVP}
\psi =  \tan^{-1}\left[ \frac{\hat{\Vec{L}} \cdot \hat{\Vec{Z}}_D -(\hat{\Vec{L}} \cdot \hat{\Vec{N}})(\hat{\Vec{Z}}_D \cdot \hat{\Vec{N}})}{\hat{\Vec{N}} \cdot (\hat{\Vec{L}} \times \hat{\Vec{Z}}_D)} \right] \,,
\end{equation}
and 
\begin{equation} \label{eq: takahasi amplitude_LVP}
A = \sqrt{\frac{5}{96}} \frac{1}{D_L}\frac{\mathcal{M}^{5/6}}{ \pi^{2/3}}\,,
\end{equation}
where $D_{L}$ is the luminosity distance of the GW source and $\mathcal{M}$ is its chirp mass.
For a binary with total mass $M = m_1 + m_2$ and mass ratio $q = m_2/m_1 \leq 1$, the symmetric mass ratio is $\eta \equiv q/(1+q)^2$ and the chirp mass is $\mathcal{M} \equiv \eta^{3/5}M$.

The spin-independent GW phase at 2PN order is \cite{PhysRevD.52.848}
\begin{align}
\label{eq: phase 2PN_LVP}
\Psi(f) &= 2 \pi f t_c - \phi_c - \frac{\pi}{4} + \frac{3}{128} \eta^{-1} x^{-5/2} \notag \\
&\qquad \times \left[ 1 + \frac{20}{9} \left(\frac{743}{336} + \frac{11}{4}\eta \right) x - 16\pi x^{3/2} + 10\Gamma x^2 \right]\,,
\end{align}
where $x \equiv (\pi Mf)^{2/3}$ is the traditional PN parameter, $t_c$ and $\phi_c$ are the time and GW phase at binary coalescence, and
\begin{equation}
\Gamma \equiv \frac{3058673}{1016064}+\frac{5429}{1008}\eta+\frac{617}{114}\eta^2 \,.
\end{equation}
The polarization phase $\phi_p$ in the strain of \cref{eq: hf A and phase_LVP} is given as
\begin{align} 
\phi_p &= \tan^{-1}\left[ \frac{2(\hat{\Vec{L}} \cdot \hat{\Vec{N}}) \tan(2\psi + \alpha)}{1 +(\hat{\Vec{L}} \cdot \hat{\Vec{N}})^2} \right]\,, \label{eq: phip_LVP}
\end{align}
and $\delta\Phi$ is an additional contribution to the GW phase for precessing BBH systems \cite{Apostolatos1994} given by integrating
\begin{equation} \label{eq: delta correction_LVP}
\frac{d\delta\Phi}{df} = \left[ \frac{\hat{\Vec{L}} \cdot \hat{\Vec{N}}}{1 -(\hat{\Vec{L}} \cdot \hat{\Vec{N}})^{2}} \right] (\hat{\Vec{L}} \times \hat{\Vec{N}}) \cdot \frac{d\hat{\mathbf{L}}}{df}\,.
\end{equation}
As we are only interested in the inspiral portion of the waveform, we cut off our waveforms at a GW frequency
\begin{equation} \label{eq: fcut_LVP}
\fcut = \frac{1}{6^{3/2}\pi M_z} = 4.3 \times 10^3\,\mathrm{Hz} \left( \frac{M_z}{M_\odot} \right)^{-1}
\end{equation}
equivalent to the quadrupole frequency at the innermost stable circular orbit of a non-spinning BH of mass $M_z$ \cite{1992ApJ...400..175B, CutlerFlanagan1994}.
Here $M_z = (1+z)M$ is the redshifted total mass of the binary.

\subsection{Gravitational lensing}
\label{subsec: lensing background}

\begin{figure}
    \centering
    \includegraphics[width=0.8\columnwidth]{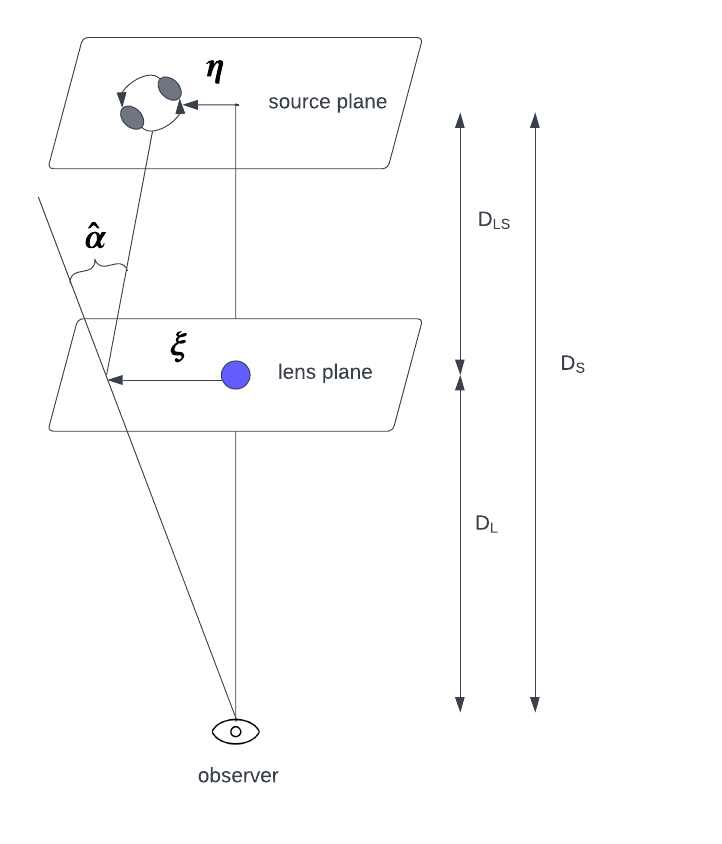}
    \caption{Schematic of a gravitational lens system, modified from Fig.~1 in \cite{Ali2023}.
    $\mathcal{D}_L$, $\mathcal{D}_{LS}$, and $\mathcal{D}_S$ are the angular-diameter distances from observer to lens, lens to source, and observer to source, respectively, while $\boldsymbol{\eta}$, $\boldsymbol{\xi}$, and $\hat{\boldsymbol{\alpha}}$ are the position in the source plane, the impact parameter in the lens plane, and the deflection angle.
    }
    \label{fig: Ali Lensing}
\end{figure}

GWs can be lensed by massive objects along the line of sight as shown in \cref{fig: Ali Lensing}, resulting in magnification and modulation of the GWs, as well as potential multiple images.
The images arrive at the detector with a time delay $\tdel$ between them due to lensing, which is a function of the geometry of the path traveled and the Shapiro time delay due to the gravitational potential of the lens.
The time delay with respect to a trajectory along the optic axis is
\begin{equation}
    t_d(\mathbf{x}, \mathbf{y}) = \frac{\mathcal{D}_S \xi_0^2 (1+z_L)}{\mathcal{D}_L \mathcal{D}_{LS}} \left[ \frac{1}{2}|\mathbf{x} - \mathbf{y}|^2 - \psi(\mathbf{x}) + \phi_m(\mathbf{y}) \right]\,,
\end{equation}
where $\mathbf{x} = \boldsymbol{\xi}/\xi_0$, $\mathbf{y} = \boldsymbol{\eta}\mathcal{D}_L / \xi_0 \mathcal{D}_S$, $\psi(\mathbf{x})$ is the lensing potential, and $\phi_m(\mathbf{y})$ is an offset term chosen such that $\min_{\mathbf{x}} (t_d) = 0$.
Here $\xi_0$ is a model-dependent characteristic length scale on the lens plane called the Einstein radius, $z_L$ is the lens redshift, $\mathcal{D}_L$, $\mathcal{D}_{LS}$, and $\mathcal{D}_S$ are the angular-diameter distances from observer to lens, lens to source, and observer to source, $\xi$ is the impact parameter in the lens plane, and $\eta$ is the location of the source with respect to the optic axis in the source plane.

The lensing amplification factor $F(f)=\hL(f)/\tilde{h}(f)$ is given by Kirchhoff's diffraction integral~\cite{TakahashiNakamura2003, Ali2023}:
\begin{equation}
    F(f) = \frac{\mathcal{D}_S \xi_0^2 (1+z_L)}{\mathcal{D}_L \mathcal{D}_{LS}} \frac{f}{i} \int d^2\mathbf{x} \exp{[2\pi i f t_{\mathrm{d}}(\mathbf{x},\mathbf{y})]},
\end{equation}
which is an integral over the lens plane, accounting for all the possible trajectories along which the wave can propagate.

\subsubsection{Geometrical-optics approximation}
\label{subsubsec: Geo-optics}

In the geometrical-optics regime in which the GW wavelength is not significantly larger than the Schwarzschild radius $R_S = 2M_L$ of the lens, discrete images form at the stationary points $x_j$ where $\nabla_\mathbf{x} t_d(\mathbf{x}, \mathbf{y}) = 0$, and only these points contribute to the lensing amplification factor~\cite{Ali2023}:
\begin{equation} \label{eq: amp factor_LVP}
    F(f) = \sum_j |\mu_j|^{1/2} \exp{\left( 2\pi if t_d(\mathbf{x}_j, \mathbf{y}) - i\pi n_j \right)},
\end{equation}
where $\mu_j = 1/\det(\delta \mathbf{y} / \delta \mathbf{x}_j)$ is the magnification of the $j$-th image and $n_j$ is the Morse index, which has the value of 0, 1/2, or 1 when the $j$-th image is a minimum, saddle point, or maximum point, respectively, of the time-delay surface.

We consider axisymmetric lens models, which result in at most two images, for our lensing analysis.
For cuspy lens models like the point mass and singular isothermal sphere, the first image is formed at a minimum of the time-delay surface and the second image is formed at the saddle point.
We parameterize the amplification factor for such two-image lenses using model-independent image parameters: the flux ratio $I = \vert \mu_-\vert/ \vert \mu_+ \vert$ and the time delay $\tdel$ between the two images, where $+$ and $-$ denote the minimum and saddle point images, respectively \cite{Ali2023}.
For a given lens model, these parameters can be inverted to obtain the source position $y$ with respect to the optical axis and the lens mass $M_L$ in the geometrical-optics approximation.
The amplification factor $F(f)$ for two-image lenses can be written in terms of time delay and flux ratio as:
\begin{equation} \label{eq: amp factor model independent_LVP}
F(f) = |\mu_{+}|^{1/2} (1 - i I^{1/2} e^{2\pi if \tdel}),
\end{equation}
The signal-to-noise ratio (SNR) of a lensed GW event is proportional to the overall normalization $|\mu_+|^{1/2}$, but the mismatch between a lensed and unlensed waveform is independent of its value.

\subsubsection{Lensed waveforms}
\label{subsubsec: lensed waveforms}

When the time delay between the two images is shorter than the time that the signal spends in the sensitivity band of a GW detector, interference between the two images will occur and induce modulations in the observed waveform.
This interference pattern is encoded by the amplification factor, and the lensed waveform is given by
\begin{align}
    \hL(f) = F(f) \hUL(f) \,, \label{eq: lensed strain_LVP}
\end{align}
where $\hUL(f)$ is the unlensed frequency-domain strain for the inspiral phase given by \cref{eq: hf A and phase_LVP}.

\begin{figure*}[t]
    \centering
    \includegraphics[width=\textwidth]{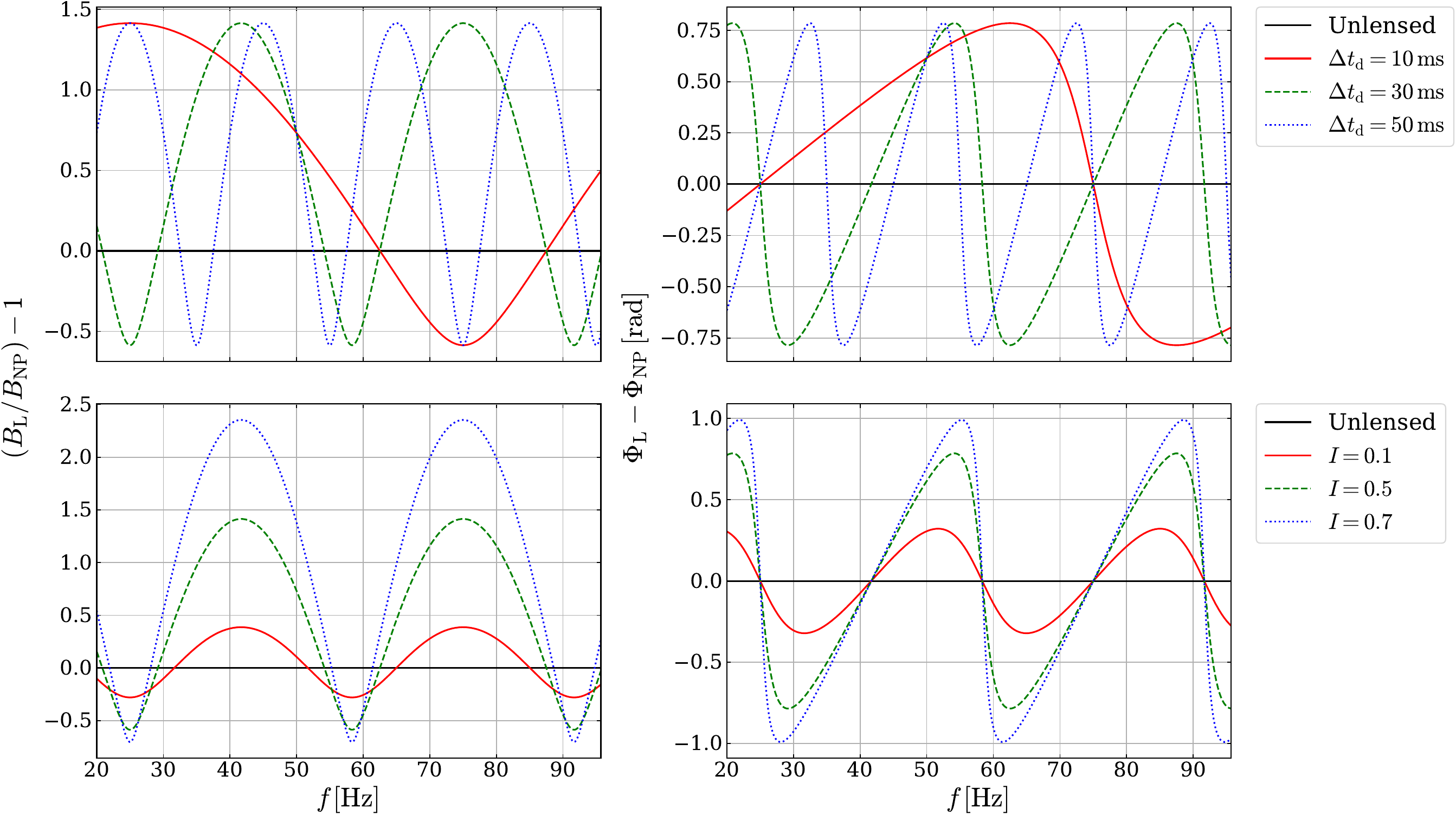}
    \caption{The amplitude ratio $\BL/\BNP$ (left panels) and phase difference $\PhiL - \PhiNP$ (right panels) of lensed (L) and non-precessing (NP) waveforms as a function of GW frequency $f$ generated by equal-mass BBHs with source-frame chirp mass of $10\,M_\odot$ and redshift $z=1$.
    The top panels fix the flux ratio $I$ of the lensed waveforms at 0.5 while varying their time delay $\tdel$; the solid red, dashed green, and dotted blue curves correspond to $\tdel = 10$, 30, and $50\,\mathrm{ms}$.
    The bottom panels fix $\tdel = 30\,\mathrm{ms}$, while the solid red, dashed green, and dotted blue curves correspond to $I = 0.1$, 0.5, and 0.7.}
    \label{fig: waveforms lensed}
\end{figure*}

The frequency-domain strain amplitude and phase of an unlensed NP inspiral waveform follow the spin-independent power law $f^{-7/6}$ and phase given in \cref{eq: hf amplitude_LVP,eq: phase 2PN_LVP}, respectively.
For a lensed waveform, the amplitude and phase modulations encoded by the lensing amplification factor of \cref{eq: amp factor model independent_LVP} are shown in \cref{fig: waveforms lensed}.
The waveforms in this figure originate from equal-mass BBHs with a chirp mass of $10\,M_\odot$ at redshift $z=1$, 
with total-angular-momentum direction $\hat{\Vec{J}}$ and sky location $\hat{\Vec{N}}$ given by $\theta_J = \pi/2$, $\Phi_J = \pi/2$, $\theta_S = \pi/4$, and $\Phi_S = 0$.
This corresponds to System 2 in \citeauthor{TamanRP2025}~\cite{TamanRP2025} which was chosen to have an edge-on orientation ($\cos\iotaJN = 0$).
Since the amplification factor is independent of orientations and sky locations, this choice is irrelevant for lensing.

The upper panels of \cref{fig: waveforms lensed} show that longer time delays $\tdel$ lead to more closely spaced interference fringes in the frequency domain, as follows from \cref{eq: amp factor model independent_LVP}.
In the lower panels of \cref{fig: waveforms lensed}, an increase in flux ratio $I$ leads to higher-amplitude interference fringes, as is also evident from \cref{eq: amp factor model independent_LVP}.
The number of interference fringes in band is independent of the flux ratio.
Lensing induces purely oscillatory changes to the GW phase $\PhiL$, which will become important later.
For a given lensed waveform, the frequency interval between consecutive fringes is the inverse of the time delay and remains constant during the inspiral because of the fixed geometry of the path between the source, the lens, and the detector.

\subsection{Precession}
\label{subsec: Precession background}

GWs from BBH inspirals are also affected by the black-hole spins and orbital angular momentum, which, when misaligned, cause precession and nutation of the orbit.
These manifest as modulations in GW amplitudes and phases~\cite{Apostolatos1994, GangardtSteinle2021, TamanRP2025}.
\citeauthor{GangardtSteinle2021}~\cite{GangardtSteinle2021} proposed five phenomenological parameters to describe generic precession, among which the precession amplitude $\langle \thLJ \rangle$ and precession frequency $\langle \OmLJ \rangle$ are sufficient to characterize regular precession.
The averaged precession parameters $\langle \thLJ \rangle$ and $\langle \OmLJ \rangle$ remain constant on the precession timescale, but vary on the radiation-reaction timescale as the gravitational waves carry away the orbital energy and angular momentum \cite{Multitimescale2015, GangardtSteinle2021, TamanRP2025}.
In regular precession (simple precession without nutation), the total angular momentum has a nearly constant direction, and the orbital angular momentum precesses on a cone whose opening angle and frequency slowly increase on the radiation-reaction timescale \cite{Multitimescale2015, GangardtSteinle2021, Apostolatos1994, TamanRP2025}.
This behavior is a special case of simple precession that occurs in binaries with single non-zero spin, precisely equal masses, or those trapped in spin-orbit resonances \cite{GangardtSteinle2021, Apostolatos1994, TamanRP2025}.
In the approximation that the averaged precession parameters $\langle \thLJ \rangle$ and $\langle \OmLJ \rangle$ retain their lowest PN order frequency dependence all the way up to $\fcut$, they can be expressed in terms of constant dimensionless precession parameters $\thtil$ and $\Omtil$ as in Eqs.~(18a) and (18b) of \citeauthor{TamanRP2025}~\cite{TamanRP2025}:
\begin{subequations} \label{E:PPfreqdep}
\begin{align} 
\langle \thLJ \rangle &= \frac{0.1\thtil}{4 \eta}\left(\frac{f}{\fcut}\right)^{1/3}\,, \label{eq: thetaLJ_LVP} \\
\langle\OmLJ \rangle &= 10^3\,\mathrm{Hz}~\Omtil \left(\frac{f}{\fcut}\right)^{5/3}\left(\frac{M}{M_\odot}\right)^{-1}. \label{eq: Omega_LJ_LVP}
\end{align}
\end{subequations}

Characterizing regularly precessing binaries requires one further parameter beyond $\thtil$ and $\Omtil$: the initial precession phase $\gammaP$, which is the value of $\PhiLJ$ when the binary enters the sensitivity band at $f = \fmin$~\cite{TamanRP2025}.
$\PhiLJ$ is the azimuthal angle of the projection of $\hat{\Vec{L}}$ onto the plane perpendicular to $\hat{\Vec{J}}$, measured from the line of nodes where that plane intersects the plane perpendicular to $\hat{\Vec{N}}$.
$\gammaP$ is a nuisance parameter that reflects the binary orientation at $f = \fmin$ and does not have any implications beyond shifting the precessional phase, as seen in \cref{eq: PhiLJ_LVP} below \cite{TamanRP2025}.

Given the dimensionless precession parameters $\thtil$ and $\Omtil$ (along with $\gammaP$), we can describe the frequency-dependent motion of the orbital angular momentum and obtain the dot product $\hat{\Vec{L}} \cdot \hat{\Vec{N}}$ for \cref{eq: hf amplitude_LVP,eq: phip_LVP,eq: delta correction_LVP}, which can be written as~\cite{TamanRP2025}, 
\begin{equation}
    \hat{\Vec{L}} \cdot \hat{\Vec{N}} = \sin\thLJ\sin\iotaJN\sin\PhiLJ + \cos\thLJ\cos\iotaJN\,, \label{eq: LdotN_LVP}
\end{equation}
where $\iotaJN$ is the angle between the total angular momentum $\hat{\Vec{J}}$ and the sky location vector $\hat{\Vec{N}}$, $\thLJ$ is the polar angle between the orbital angular momentum $\hat{\Vec{L}}$ and the total angular momentum $\hat{\Vec{J}}$ given by \cref{eq: thetaLJ_LVP} for regular precession, and $\PhiLJ$ is the azimuthal angle for the orbital angular momentum $\hat{\Vec{L}}$ in the plane perpendicular to $\hat{\Vec{J}}$.
$\PhiLJ$ is obtained by integrating $\langle \OmLJ \rangle$ in \cref{eq: Omega_LJ_LVP}~\cite{TamanRP2025},
\begin{equation} \label{eq: PhiLJ_LVP}
\PhiLJ = \gammaP + \int^f_{\fmin} \langle\OmLJ\rangle \left( \frac{df'}{dt} \right)^{-1} df'~,
\end{equation}
where $\fmin = 20\,\mathrm{Hz}$ is the floor of the LVK sensitivity band, and to 1.5PN order, $df/dt$ is given by
\begin{equation} \label{E:df/dt_LVP}
\frac{df}{dt} = \frac{96\eta}{5\pi M^2} x^{11/2} \left[ 1 - \left( \frac{743}{336} + \frac{11}{4}\eta \right)x + 4\pi x^{3/2} \right] \,,
\end{equation}
with $x \equiv (\pi Mf)^{2/3}$.
We only use $df/dt$ at the lowest PN order in \cref{eq: PhiLJ_LVP}.

\subsubsection{Regularly precessing waveforms}
\label{subsubsec: RP waveforms}

\begin{figure*}[t]
    \centering
    \includegraphics[width=\textwidth]{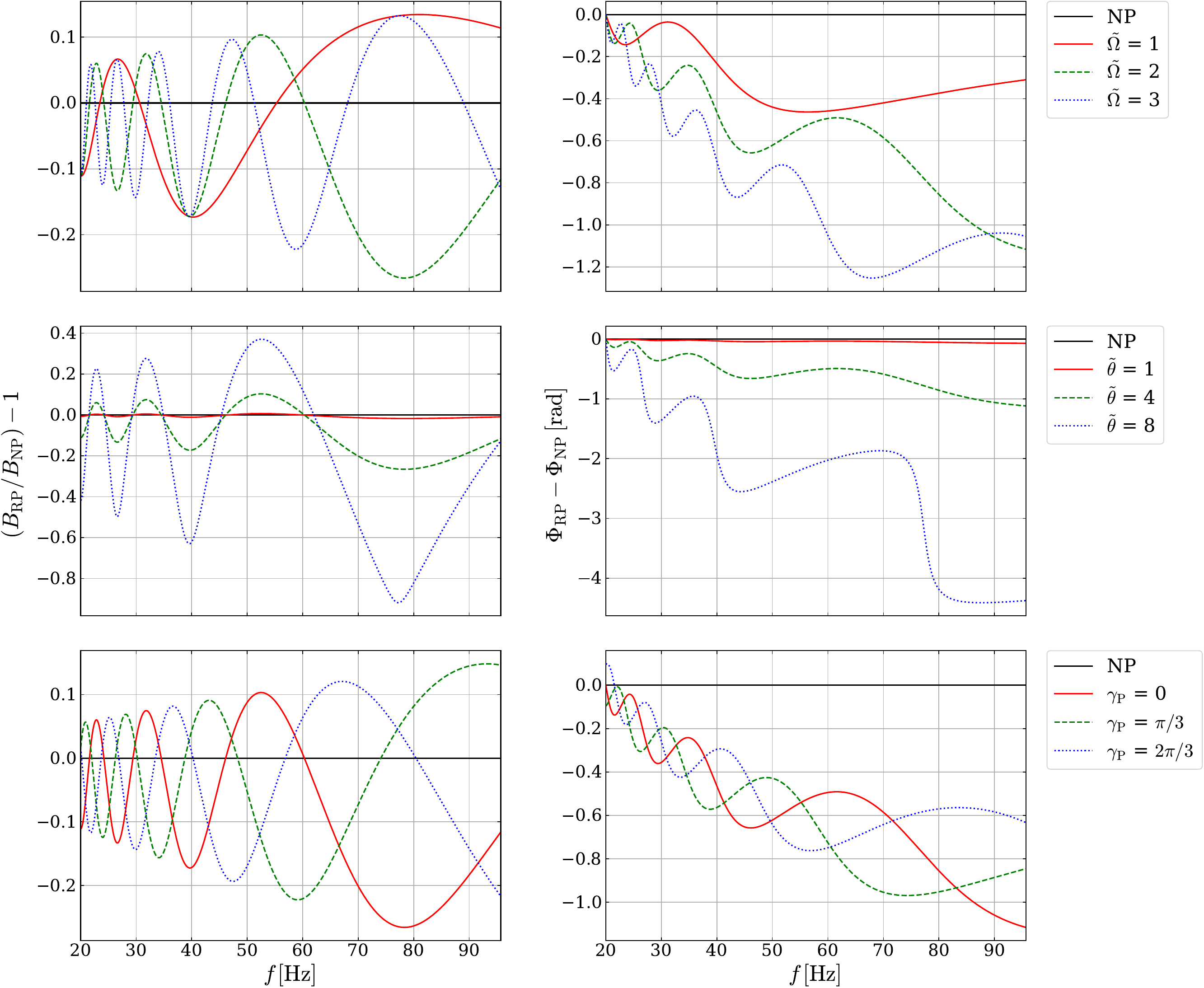}
    \caption{
    The amplitude ratio $\BRP/\BNP$ (left panels) and phase difference $\PhiRP - \PhiNP$ (right panels) of regularly precessing (RP) and non-precessing (NP) waveforms as a function of GW frequency $f$ generated by equal-mass BBHs with source-frame chirp mass of $10\,M_\odot$ and redshift $z=1$.
    The default values of the precession frequency, precession amplitude, and precession phase at $\fmin = 20\,\mathrm{Hz}$ are $\Omtil = 2$, $\thtil = 4$, and $\gammaP = 0$.
    While holding other parameters at these default values, in the top row, we vary $\Omtil = 1$ (solid red), $2$ (dashed green), and $3$ (dotted blue).
    In the middle row, we vary $\thtil = 1$ (solid red), $4$ (dashed green), and $8$ (dotted blue).
    In the bottom row, we vary $\gammaP = 0$ (solid red), $\pi/3$ (dashed green), and $2\pi/3$ (dotted blue).}
    \label{fig: waveforms RP}
\end{figure*}

Using the model for regularly precessing waveforms presented in \citeauthor{TamanRP2025}~\cite{TamanRP2025} and summarized above, we illustrate how the waveform changes with precession parameters in \cref{fig: waveforms RP}.
The waveforms again originate from equal-mass BBHs with a chirp mass of $10\,M_\odot$ with the same redshift, orientation, and sky location as \cref{fig: waveforms lensed}.
We chose these values corresponding to System 2 from~\cite{TamanRP2025} because its edge-on geometry produces the cleanest modulation patterns (see Figs.~3 and 4 in~\cite{TamanRP2025} for reference).
The default values of the precession frequency $\Omtil = 2$ and amplitude $\thtil = 4$ are close to the median values for a population of equal-mass BBHs with isotropically oriented, maximal spins.
In the top and middle rows of \cref{fig: waveforms RP}, we vary the precession frequency and amplitude between the 5\textsuperscript{th} ($\Omtil = 1$ and $\thtil = 1$) and 95\textsuperscript{th} ($\Omtil = 3$ and $\thtil = 8$) percentile values for this BBH population \cite{TamanRP2025}.

The top row of panels in \cref{fig: waveforms RP} shows that increasing $\Omtil$ leads to more precession cycles in both the strain-amplitude ratio $\BRP/\BNP$ and the phase difference $\PhiRP - \PhiNP$ between regularly precessing (RP) and non-precessing (NP) waveforms.  The middle row shows that increasing $\thtil$ gives rise to larger-amplitude modulations of both $\BRP/\BNP$ and $\PhiRP - \PhiNP$.
The lower row shows that varying $\gammaP$ merely shifts the phase of these precessional modulations.

The similarity between the waveforms shown in \cref{fig: waveforms lensed,fig: waveforms RP} suggests possible degeneracy between the signatures of lensing and precession.
Interestingly, unlike the constant spacing $\tdel^{-1}$ between the interference fringes of lensed waveforms, the frequency separation $\Delta f_{\mathrm{pre}}$ between consecutive peaks (or troughs) of the precession cycles increases with frequency as $\Delta f_{\mathrm{pre}} \propto (df/dt)t_{\mathrm{pre}} \propto f^2$.
This is because as BBHs inspiral towards merger, the ratio between the precession timescale and the radiation-reaction timescale increases (this ratio scales as $x^{-3/2}$ at lowest PN order).
In addition, the modulation caused by regular precession qualitatively differs from that of lensing in that regular precession induces both oscillatory and secular phase changes to the GW phase, while lensing only induces oscillatory phase evolution.
The distinctive contribution of precession to the secular phase may help lift the lensing-precession degeneracy.
Yet this secular contribution may be suppressed given that it increases with both the precession amplitude and frequency, as shown in the top right and middle right panels of \cref{fig: waveforms RP}.
In the next subsection, we describe the mismatch $\epsilon$ between waveforms, which we will use to quantitatively investigate the degree of lensing-precession degeneracy throughout parameter space.

\subsection{Match-filtering and mismatch}
\label{subsec: match-filtering}

To quantify differences between waveforms, we utilize the match function as implemented in the Python package \texttt{pycbc.filter} \cite{alex_nitz_2022_6324278}.
A brief review of the match statistic is provided below.

The inner product between two waveforms $\tilde{h}_1(f)$, $\tilde{h}_2(f)$ in the frequency domain is defined as
\begin{equation} \label{eq:inner_product_LVP}
    \langle h_1 | h_2 \rangle \equiv 4 \mathrm{Re} \int_{\fmin}^{\fcut} \frac{h_1(f)h_2^*(f)}{S_n(f)} df\,,
\end{equation}
where $S_n(f)$ is the power spectral density (PSD) of the detector noise~\cite{Finn1992, Wiseman1992}.

We define the signal-to-noise ratio (SNR) between the source waveform $h_{\mathrm{s}}$ and a template waveform $h_{\mathrm{t}}$ as \cite{CutlerFlanagan1994}
\begin{equation} \label{eq: SNR_LVP}
    \rho = \langle h_{\mathrm{s}} | h_{\mathrm{t}} \rangle^{1/2}\,.
\end{equation}
The match between the waveforms $h_{\mathrm{s}}$ and $h_{\mathrm{t}}$ is defined as their normalized inner product maximized over the time and phase of coalescence
\begin{equation} \label{eq:match_LVP}
    \mathrm{M}(h_{\mathrm{s}}, h_{\mathrm{t}}) \equiv \underset{t_c, \phi_c}{\max} \frac{\langle h_{\mathrm{s}} | h_{\mathrm{t}} \rangle}{\sqrt{\langle h_{\mathrm{s}} | h_{\mathrm{s}} \rangle \langle h_{\mathrm{t}} | h_{\mathrm{t}} \rangle}}\,.
\end{equation}
The mismatch is defined as
\begin{equation} \label{eq:mismatch_LVP}
    \epsilon(h_{\mathrm{s}}, h_{\mathrm{t}}) \equiv 1 - \mathrm{M}(h_{\mathrm{s}}, h_{\mathrm{t}})\,.
\end{equation}
For detector noise, we use the PSD for Advanced LIGO (aLIGO) \cite{Satyaprakash2009}.
When comparing RP templates with lensed sources, we minimize the mismatch over the precessional phase $\gammaP$ since it is a nuisance parameter,
\begin{equation} \label{E:epsP}
   \epsP = \min\limits_{\gammaP} \epsilon (h_{\mathrm{s}}, h_{\mathrm{t}})\,.
\end{equation}

The Lindblom distinguishability criterion~\cite{Lindblom2008} establishes the condition under which a source waveform $\hS$ can be distinguished from a template $\hT$:
\begin{equation} \label{eq: Lindblom_LVP}
    \epsST \geq \frac{1}{2\rho_{\mathrm{s}}^2} \,,
\end{equation}
where $\rho_{\mathrm{s}}$ is the source SNR.
Equivalently, the SNR threshold for distinguishing two waveforms with mismatch $\epsST$ scales as $\rho_{\mathrm{min}} \propto \epsST^{-1/2}$.

\section{Results}
\label{sec: Results}

\subsection{Lensing-precession mismatch vs. precession amplitude and frequency}
\label{subsec: RP templates and a lensed source}

To establish degeneracies and differences between lensing-induced and precession-induced modulations of GW waveforms, we compare a lensed, NP source against unlensed RP templates.
We hold all source parameters---including sky location, orientation, redshift, chirp mass, and mass ratio---fixed between the source and the templates.
Our three-dimensional template bank is parameterized by the three precessional parameters $\Omtil$, $\thtil$, and $\gammaP$.
Given a lensed source with fixed flux ratio $I$ and time delay $\tdel$ between the images, we explore this RP parameter space to see whether an RP template yields a better match (lower mismatch) than an NP template with the same NP source parameters.

\begin{figure*}
    \centering
    \includegraphics[width=\textwidth]{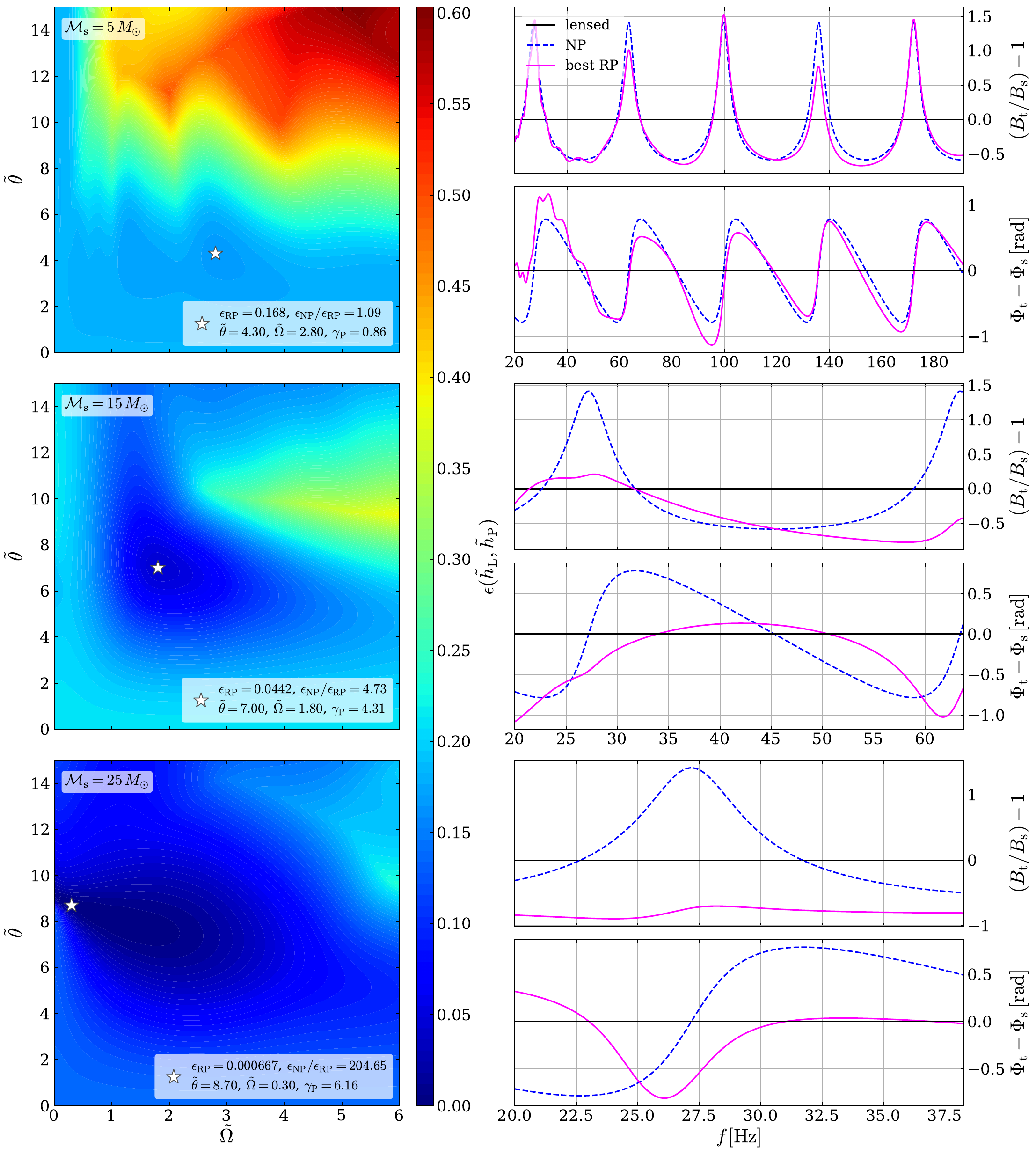}
    \caption{
    \textbf{Left:} Contour plots of the mismatch $\epsP$ as a function of the dimensionless precession frequency $\Omtil$ and amplitude $\thtil$.
    The source waveforms are lensed BBHs with flux ratio $I = 0.5$ and time delay $\tdel = 30\,\mathrm{ms}$ at redshift $z=1$.
    The sky location and system orientation are those of System 2 listed in \cref{tab: source parameters_Lensed} below.
    The top, middle, and bottom rows correspond to source-frame chirp masses $\Mcs = 5$, $15$, and $25\,M_\odot$.
    Stars mark the global minima in each plot and the legends indicate the minimum mismatch $\epsRP$, the ratio $\epsNP/\epsRP$ by which this mismatch is reduced compared to that with an NP template, and the values of the precession parameters characterizing the best-fitting RP template.
    \textbf{Right:} Amplitude ratios $\BT/\BS$ and phase differences between the lensed source waveforms and the NP (dashed blue) and best-fitting RP (solid magenta) templates as functions of GW frequency $f$.
    The top, middle, and bottom rows correspond to the same chirp masses as the left panels. 
    }
    \label{fig:sys2_combined_contour_waveform}
\end{figure*}

\done{\mk{For the publication, we should have nicer increments on the left contour plot.
Also add red points at the actual minima.}
\ts{I agree!}}

\Cref{fig:sys2_combined_contour_waveform} illustrates the mismatch between a lensed source and precessing templates for three representative source chirp masses $\Mcs \in \{5, 15, 25\}\,M_\odot$.
The source is lensed by an axisymmetric lens to produce two images with a flux ratio of $I = 0.5$ and a time delay of $\tdel = 30\,\mathrm{ms}$ between them.
Each contour plot is computed by scanning $\Omtil \in [0, 6]$ and $\thtil \in [0, 15]$, with $\gammaP$ optimized over $[0, 2\pi]$ at each grid point to minimize the mismatch.
The starred points mark the global minima, and the legends quote the corresponding best-fit parameters.
Although typical populations of equal-mass binaries with maximal spins and isotropic orientations exhibit smaller precession parameters, with 95\textsuperscript{th} percentiles of $\thtil \approx 8.05$ and $\Omtil \approx 2.57$~\cite{TamanRP2025}, we expand the parameter space to $\thtil \in [0, 15]$ and $\Omtil \in [0, 6]$ to systematically map the trends and behavior of the degeneracy across diverse lensing configurations. 

The bottom left corner of each contour plot, where $\Omtil = \thtil = 0$, corresponds to the NP limit.
In all three rows, the global minimum lies away from this corner, demonstrating that an RP template outperforms an NP one for these lensed sources regardless of chirp mass.
The location of the minimum indicates the precession frequency and amplitude of the RP waveform that best replicate the lensing-induced modulations in the amplitude and phase of the source waveform.

The source chirp mass sets the cutoff frequency $\fcut \propto \Mcs^{-1}$ by \cref{eq: fcut_LVP} and hence the number of interference fringes in band. 
For a lensed source with time delay $\tdel$, this number is
\begin{align}
    \Nfringe &= \tdel \times (\fcut - \fmin)\,, \notag \\ 
    &= \tdel \left(\frac{\eta^{3/5}}{6^{3/2}\pi \Mcs(1+z)} - \fmin\right)\,,\label{eq: number of lensing modulations}
\end{align}
where $\eta$ is the source symmetric mass ratio, $\Mcs$ is the source chirp mass, $z$ is the redshift, and $\fmin$ (= 20 Hz for aLIGO) is the sensitivity floor of the detector.
For small chirp masses (large $\Nfringe$), RP templates with frequency-dependent spacing $\Delta f_{\mathrm{pre}} \propto f^2$ between successive crests and troughs struggle to match the uniform spacing $\Delta f_{\mathrm{lens}} = \tdel^{-1}$ between interference fringes in lensed waveforms.

For $\Mcs = 5\,M_\odot$, shown in the top row of \cref{fig:sys2_combined_contour_waveform}, $\Nfringe = 5.06$ and precession is incapable of mimicking the effects of lensing.
Precession only reduces the mismatch by a factor $\epsNP/\epsRP = 1.09$ for the best-fitting RP template, and there are several alternative local minima with similar values.
The right panel shows that the residuals for the best-fitting RP template (solid magenta curves) are almost as large as those for the NP template (dashed blue curves).
The fact that precession also provides a secular contribution to the GW phase as shown in the right panels of \cref{fig: waveforms RP} also prevents a good match between the long lensed and RP waveforms at such a low chirp mass.

As the chirp mass increases to $\Mcs = 15\,M_\odot$, shown in the middle row of \cref{fig:sys2_combined_contour_waveform}, the waveform shortens due to the reduced value of $\fcut$ and $\Nfringe$ decreases to $1.28$ according to 
\cref{eq: number of lensing modulations}.
As seen in the left panel, a unique, well-defined minimum in the mismatch emerges with $\epsNP/\epsRP = 4.31$, and the interference fringes are reasonably well matched by precessional modulations as can be seen by the reduced residuals compared to the NP template shown in the right panel.

Finally, for $\Mcs = 25\,M_\odot$ shown in the bottom row, the minimum in the mismatch contour plot has broadened and deepened.
With only $\Nfringe = 0.53$ interference fringes in band, excellent matching RP templates exist in much of the parameter space with a global minimum of $\epsNP/\epsRP = 204.65$.
The best-fitting RP template can largely suppress the amplitude and phase residuals with the lensed source waveform as can be seen in the right panel.

The Lindblom criterion of \cref{eq: Lindblom_LVP} provides an estimate of the SNR needed to distinguish two waveforms.
Since the SNR is inversely proportional to the luminosity distance $D$ of the source, the horizon for identifying lensing is reduced by a factor $D_{RP}/D_{NP} = (\epsNP/\epsRP)^{-1/2}$ when precessing templates are included in the search.
As shown above, this factor can be as small as $D_{RP}/D_{NP} \lesssim 0.07$ for $\Mcs \gtrsim 25\,M_\odot$ at $z=1$.

\subsection{Lensing-precession mismatch vs. chirp mass and time delay}

\label{subsec: NP vs RP}

Unlensed NP templates never produce the oscillatory modulations in the amplitude and phase of the waveform that are expected from lensed sources as seen in \cref{fig: waveforms lensed}.
The time delay $\tdel$ between the two images and the source chirp mass $\Mcs$ set the number of interference fringes $\Nfringe$ observed in band by \cref{eq: number of lensing modulations}.
For $\Nfringe > 1$, significant mismatches are expected between lensed sources and NP templates; \cref{appendix sec: analytical mismatch} investigates this mismatch analytically.
We hypothesize that the oscillatory features of RP templates can mimic the interference fringes of lensed waveforms for $\Nfringe \lesssim 3$, but for higher values of $\Nfringe$, the non-uniform spacing between the precessional oscillations and the secular precessional contributions to the GW phase largely break the lensing-precession degeneracy.

\begin{figure*}
    \centering
    \includegraphics[width=1\textwidth]{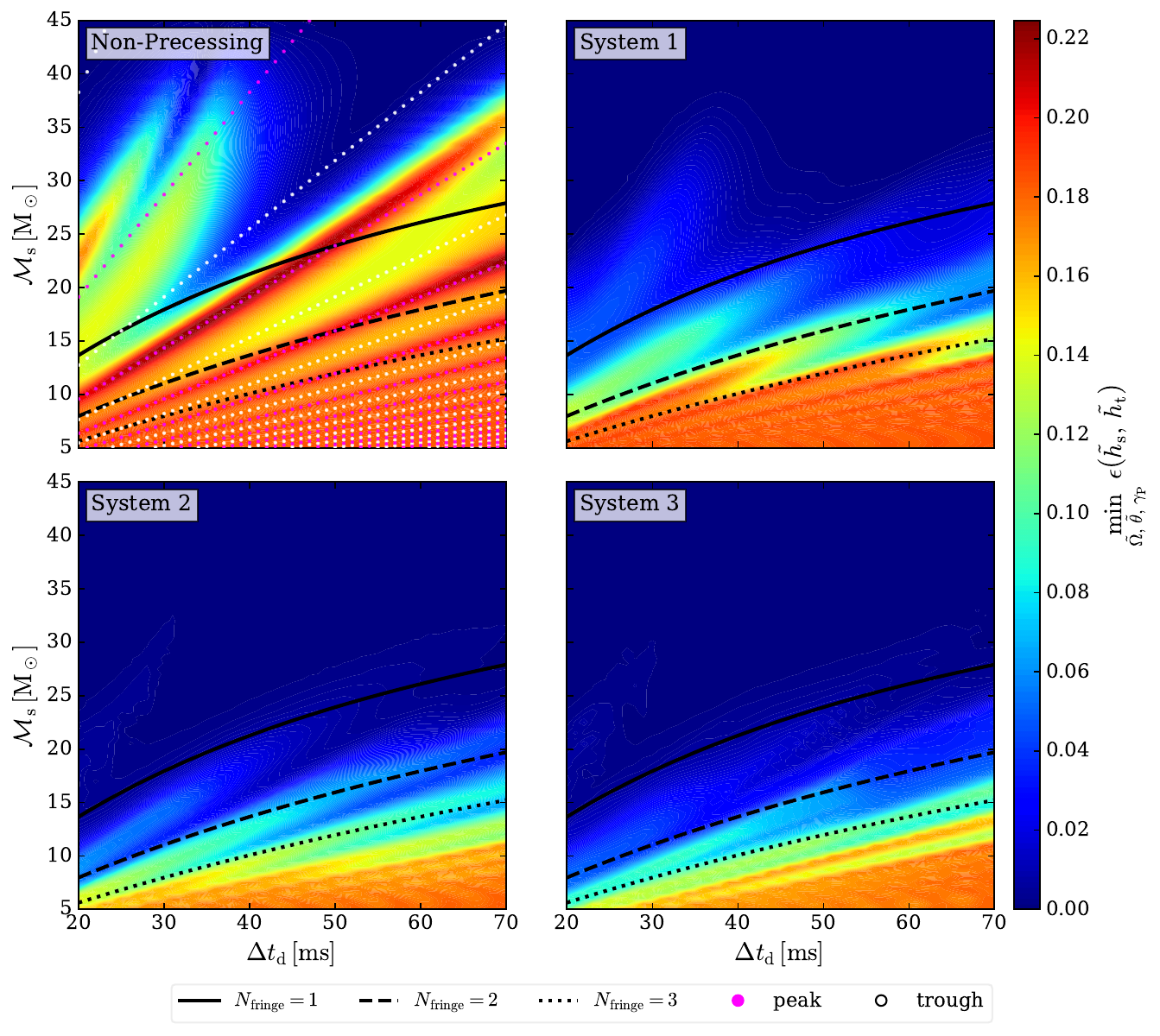}
    \caption{
        Mismatch $\epsilon$ for lensed BBH sources as a function of lensing time delay $\tdel$ and source chirp mass $\Mcs$, with fixed flux ratio $I=0.5$ and source redshift $z=1$.
        The top left panel shows the mismatch with NP templates, while the top right, bottom left, and bottom right panels show the mismatches with RP templates with the direction of the total angular momentum $\mathbf{J}$ corresponding to Systems 1, 2, and 3 specified in \cref{tab: source parameters_Lensed}.
        These mismatches are minimized over the precessional amplitude $\thtil$, frequency $\Omtil$, and initial phase $\gammaP$.
        The solid, dashed, and dotted black curves correspond to $\Nfringe = 1$, $2$, and $3$ in band.
        In the top left panel, the dotted magenta (white) lines correspond to analytic predictions of the maxima (minima) of the mismatch between lensed waveforms and NP templates given by \cref{eq: mismatch extrema mcz}.  
    }
    \label{fig: compare LvsNP LvsRP td mcz}
\end{figure*}

\begin{table}
    \centering
    \begin{tabular}{ |c|c|c|c|  }
    \hline
     Parameter & System 1 & System 2 & System 3 \\
     \hline
     $\theta_J$ & $\pi/4$ & $\pi/2$ & $8\pi/9$ \\ \hline
     $\Phi_J$ & 0 & $\pi/2$ & $\pi/4$ \\ \hline
     $\cos\iotaJN$ & 1 & 0 & -0.493 \\ \hline
    \end{tabular}
    \caption{Spherical coordinates $\theta_J, \Phi_J$ specifying the direction of the total angular momentum $\mathbf{J}$ in the detector frame (from~\cite{TamanRP2025}), and the inclination $\iotaJN$ between $\hat{\mathbf{J}}$ and the sky location $\hat{\mathbf{N}}$.
    All three BBH systems have a fixed sky location of $\theta_S = \pi/4, \Phi_S = 0$.
    }
    \label{tab: source parameters_Lensed}
 \end{table}

We investigate the validity of this hypothesis numerically in \cref{fig: compare LvsNP LvsRP td mcz}.
The top left panel is a contour plot of the mismatch $\epsilon$ between lensed sources and unlensed NP templates as a function of the lensing time delay $\tdel$ and chirp mass $\Mcs$ of the source.
Although we used System~2 to prepare this panel, it is independent of the direction of the total angular momentum $\mathbf{J}$ since its effect on the GW amplitude cancels in the numerator and denominator of the match in \cref{eq:match_LVP}.
We see oscillatory features related to the interference features of the lensed source waveform; the locations of these features are quite accurately predicted by \cref{eq: mismatch extrema mcz}.
The portion of the plot below the solid black curve has $\Nfringe > 1$ and generally has large mismatches $\epsilon \approx 1 - (1+I)^{-1/2} = 0.18$ consistent with the $\Nfringe \to \infty$ prediction of \citeauthor{Ali2023}~\cite{Ali2023} reviewed at the end of \cref{appendix sec: analytical mismatch}.

Above this solid black curve, where $\Nfringe < 1$, excellent matches are possible even with NP templates.
There are exceptions, however, particularly near the magenta lines where peaks of $\epsilon$ are predicted.

The top right, bottom left, and bottom right panels of \cref{fig: compare LvsNP LvsRP td mcz} show the mismatches between the same lensed source waveforms and RP templates with the direction of the total angular momentum $\mathbf{J}$ given by Systems 1, 2, and 3 listed in \cref{tab: source parameters_Lensed}.
These mismatches are minimized with respect to the precessional amplitude $\thtil$, frequency $\Omtil$, and initial phase $\gammaP$.
The solid, dashed, and dotted black curves correspond to $\Nfringe = 1$, $2$, and $3$ and divide the plots into three regions:

\begin{itemize}
\item Region I ($\Nfringe < 1$, above solid black curve): strong lensing-precession degeneracy characterized by tiny mismatches between lensed source waveforms and RP templates with a broad range of precession amplitude $\thtil$ and frequency $\Omtil$, as in the bottom left panel of \cref{fig:sys2_combined_contour_waveform};

\item Region II ($1 < \Nfringe < 3$, between solid and dotted black curves): significant lensing-precession degeneracy characterized by $\epsNP/\epsRP - 1 \sim \mathcal{O}(1)$ for unique values of $\thtil$ and $\Omtil$, as in the middle left panel of \cref{fig:sys2_combined_contour_waveform};

\item Region III ($\Nfringe > 3$, below dotted black curve): minimal lensing-precession degeneracy characterized by $\epsNP/\epsRP - 1 \lesssim \mathcal{O}(0.1)$ for several shallow minima in the $\thtil$--$\Omtil$ plane, as in the top left panel of \cref{fig:sys2_combined_contour_waveform}.

\end{itemize}
Even with RP templates, the mismatch $\epsilon \approx 1 - (1+I)^{-1/2} = 0.18$ in the bottom right corners of these contour plots illustrates that the lensing-precession degeneracy is fully broken in the $\Nfringe \to \infty$ limit.

The three regions defined above largely bear out our hypotheses about the lensing-precession degeneracy.
In Region I, the lensed source waveforms are generally short, and the absence of multiple interference fringes implies that the lensed source waveforms can be well-fit by RP templates with small values of $\Omtil$ and correspondingly small secular precessional contributions to the GW phase.
The large mismatches with NP templates above the solid black curve in the top left panel of \cref{fig: compare LvsNP LvsRP td mcz} are almost entirely absent from the other panels of this figure.
System 1, the face-on case with $\hat{\mathbf{N}} \parallel \hat{\mathbf{J}}$ shown in the top right panel, is a partial exception to this result because regular precession only produces oscillatory contributions to the GW amplitude and phase for very large values of $\thtil$ for the face-on case, as can be seen in the first columns of Figs.~3 and 4 of \citeauthor{TamanRP2025}~\cite{TamanRP2025}.

Region II between the solid and dotted black curves retains a significant level of lensing-precession degeneracy.
The RP templates can often provide reasonable matches to the small number $\Nfringe < 3$ of interference fringes of the lensed source waveforms in the sensitivity band, and only a modest amount of secular phase is accumulated by the RP templates.
For the face-on System 1, the RP templates again struggle to match the lensed source waveforms, particularly for low chirp masses $\Mcs$.
The similarity between the contour plots for Systems 2 and 3 shown in the bottom panels suggests that they better reflect the typical performance of RP templates.

For Region III below the dotted black curve, the lensing-precession degeneracy is largely broken, with the RP templates unable to reduce the mismatch below $\epsilon \approx 0.1$ anywhere in the parameter space.
For such long waveforms, the secular effects of precession generically yield large GW phase shifts with respect to the lensed source waveforms, and the unevenly spaced precessional oscillations cannot match the uniformly spaced interference fringes.
The lack of oscillation in the RP templates for the face-on System 1 again leads to the worst matches with the lensed source waveforms.

\subsection{Lensing-precession mismatch vs. flux ratio and time delay}

\begin{figure*}
    \centering
    \includegraphics[width=\textwidth]{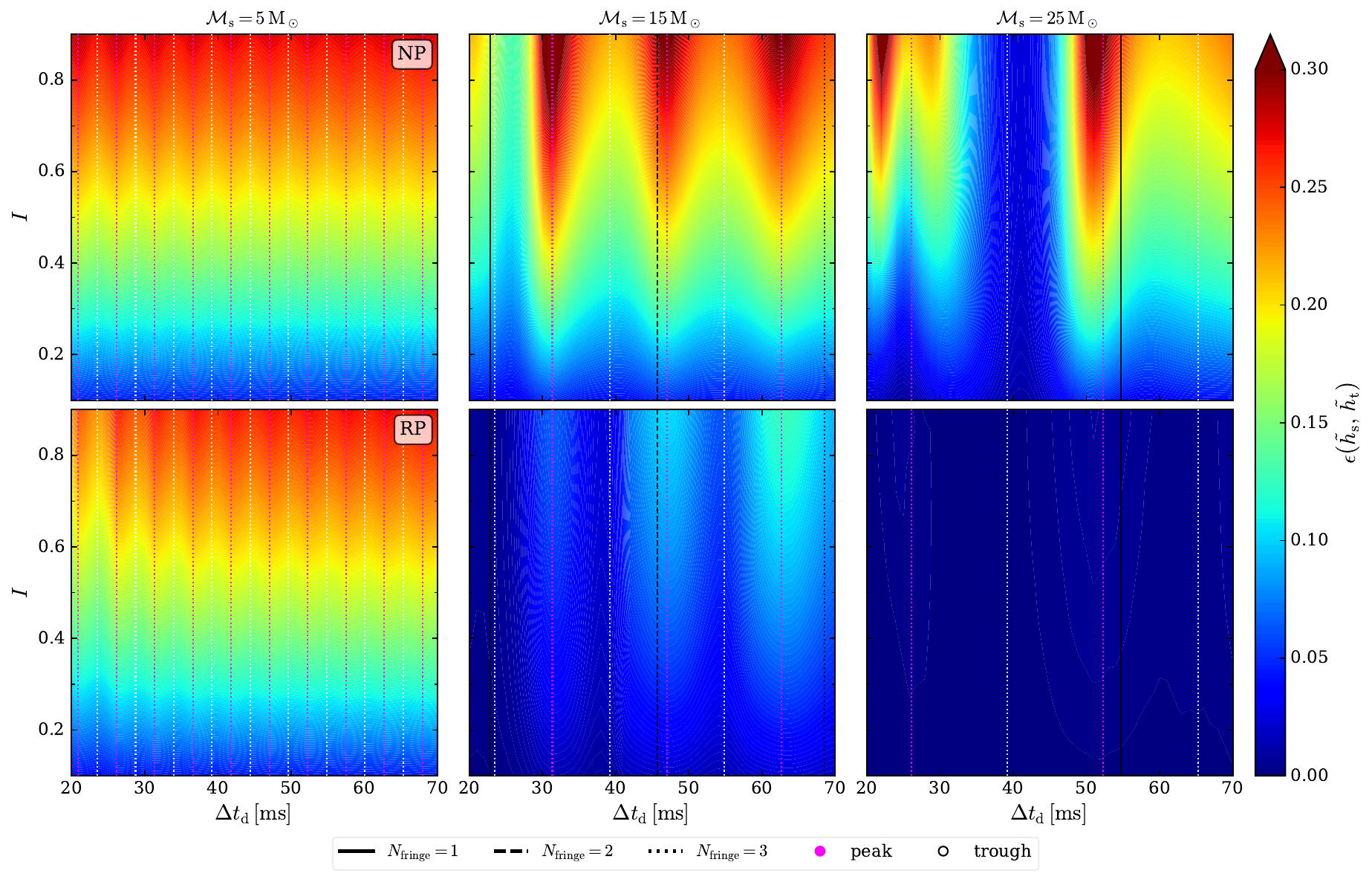}
    \caption{
    \textbf{Top row:} Contour plots of the mismatch $\epsilon$ between lensed source waveforms and NP templates as a function of the time delay $\tdel$ and flux ratio $I$ between images.
    The left, center, and right panels in both rows correspond to chirp masses $\Mcs$ of $5$, $15$, and $25\,M_\odot$.
    All plots have the sky location and binary orientation of System 2 listed in \cref{tab: source parameters_Lensed} and are at redshift $z = 1$.
    The dotted magenta (white) lines correspond to peaks (troughs) of $\epsilon$ estimated by \cref{eq: mismatch extrema mcz}, while the solid, dashed, and dotted black lines correspond to $\Nfringe = 1$, $2$, and $3$ according to \cref{eq: number of lensing modulations}.
    \textbf{Bottom row:} Contour plots of the mismatch $\epsilon$ between the same lensed source waveforms in the top row and RP templates where $\epsilon$ is minimized for each value of $\tdel$ and $I$ with respect to the precession amplitude $\thtil$, frequency $\Omtil$, and initial phase $\gammaP$.
    }
    \label{fig:compare_LvsNP_LvsRP_I_td_z1_edgeon}
\end{figure*}

\Cref{fig:compare_LvsNP_LvsRP_I_td_z1_edgeon} extends the analysis of \cref{fig: compare LvsNP LvsRP td mcz} by examining how the mismatch $\epsilon$ varies with the 
flux ratio $I$ of the lensed source waveforms.
As in \cref{fig:sys2_combined_contour_waveform}, we choose three source chirp masses, $\Mcs \in \{5, 15, 25\}\,M_\odot$, representing each of the three regions discussed above, characterized by the number $\Nfringe$ of interference fringes in the band.

The top row shows the mismatch $\epsilon$ between lensed source waveforms and NP templates.
Because both the source and template waveforms are non-precessing, $\epsilon$ depends only on the intrinsic shape of the lensing amplification factor for a two-image lens and is independent of sky location and binary orientation.
The time delay $\tdel$ controls the number of interference fringes via \cref{eq: number of lensing modulations}; a longer delay corresponds to more fringes, increasing the mismatches with unmodulated NP templates.
The flux ratio $I$ determines the amplitude of the interference fringes and sets the limit $\epsNP \to 1-(1+I)^{-1/2}$ as $\Nfringe \to \infty$ \cite{Ali2023}.
This limit rises from $0.0465$ at $I=0.1$ to $0.27$ at $I=0.9$; the right edges of the upper panels in \cref{fig:compare_LvsNP_LvsRP_I_td_z1_edgeon} are close to this limit.

In the top row, the chirp masses $\Mcs = 5\,M_\odot$ and $15\,M_\odot$ shown in the left and middle panels belong to Regions III and II respectively; both have $\Nfringe > 1$.
The mismatches exhibit well-defined peaks and troughs at values of $\tdel$ predicted by \cref{eq: mismatch extrema mcz} and marked by vertical dotted magenta and white lines.
The large $\Nfringe$ limit of $\epsNP$ approximately holds throughout the entire upper left panel ($\Mcs = 5\,M_\odot$), while this approximation breaks down in the middle panel and to a greater extent in the right panel where $\Nfringe > 1$ only for $\tdel \gtrsim 55\,\mathrm{ms}$ as shown by the vertical solid black line.

The bottom row shows the mismatch $\epsRP$ between the same lensed sources with RP templates minimized over the RP parameter space $(\Omtil, \thtil, \gammaP)$.
The mismatch is only mildly reduced for $\Mcs = 5\,M_\odot$ (Region III) consistent with the bottom edge of the bottom left panel in \cref{fig: compare LvsNP LvsRP td mcz}.
The large number of interference fringes in band for such a small chirp mass would require a large precession frequency to produce a similar number of precessional oscillations, and the associated secular phase accumulation would prevent a close match regardless of the precession amplitude.
For $\Mcs = 15\,M_\odot$, most of the panel, between the vertical solid and dotted black lines, is in Region II.
There is a dramatic reduction in the mismatch with RP templates compared to NP templates; $\epsRP \lesssim 0.1$ for all but the highest flux ratios $I \gtrsim 0.7$.
For $\Mcs = 25\,M_\odot$, the majority of the panel, left of the vertical solid black line, is in Region I.
There are uniformly low mismatches $\epsRP \lesssim 0.01$ in this panel, demonstrating the very strong lensing-precession degeneracy throughout Region I.

\begin{figure*}
    \centering
    \includegraphics[width=\textwidth]{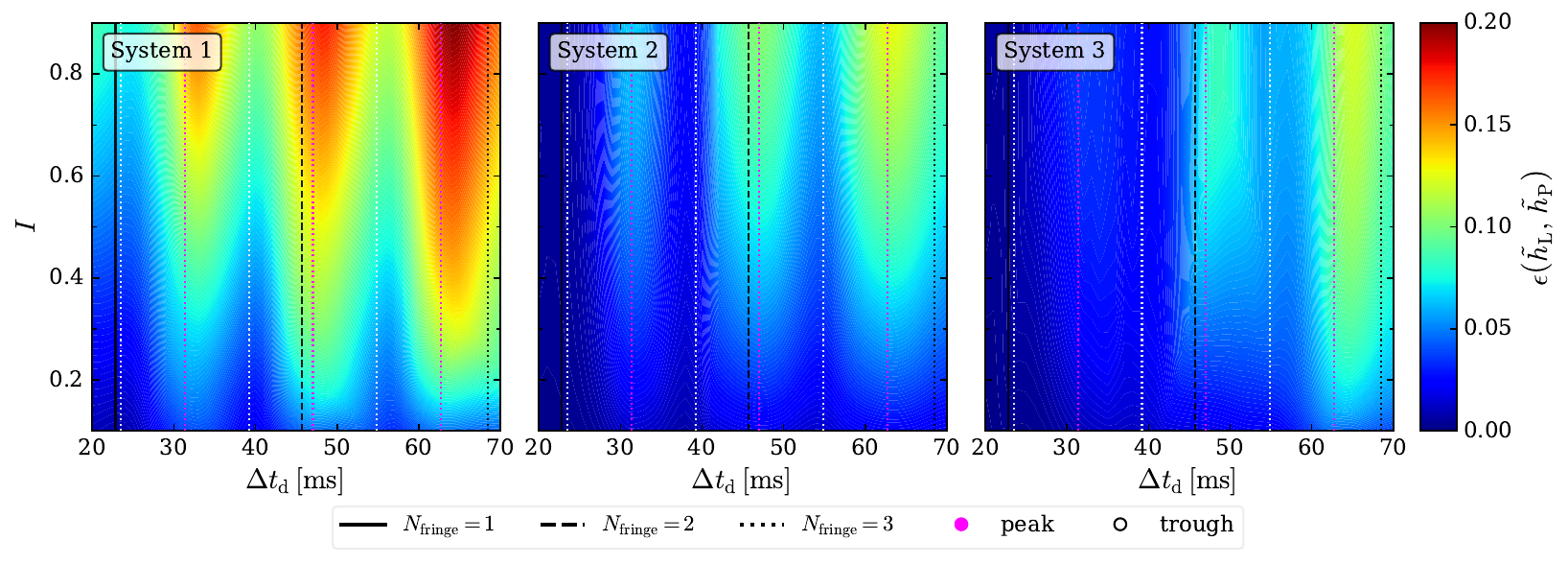}
    \caption{
    Mismatches $\epsRP$ between lensed BBH sources and RP templates as a function of lensing time delay $\tdel$ and flux ratio $I$.
    These mismatches are minimized with respect to the precession amplitude $\thtil$, frequency $\Omtil$, and initial phase $\gammaP$ of the RP templates.
    The sources have a chirp mass $\Mcs = 15\,M_\odot$ and are located at redshift $z=1$.
    The left, middle, and right panels correspond to the face-on, edge-on, and random binary orientations of Systems 1, 2, and 3 listed in \cref{tab: source parameters_Lensed}.
    The solid, dashed, and dotted black lines correspond to 1, 2, and 3 in-band interference fringes, while the dotted magenta and white lines mark the peaks and troughs of the mismatch between lensed sources and NP templates predicted by \cref{eq: mismatch extrema mcz}.
    }
    \label{fig:compare_LvsRP_I_td_z1_mcz15_allsys}
\end{figure*}

\Cref{fig:compare_LvsRP_I_td_z1_mcz15_allsys} shows the mismatch $\epsRP$ between lensed sources with chirp mass $\Mcs = 15\,M_\odot$ and RP templates minimized with respect to precession amplitude $\thtil$, frequency $\Omtil$, and initial phase $\gammaP$ for the face-on, edge-on, and random binary orientations of Systems 1, 2, and 3 listed in \cref{tab: source parameters_Lensed}.
This choice of chirp mass is the same as the middle panels of \cref{fig:compare_LvsNP_LvsRP_I_td_z1_edgeon} and primarily shows Region II ($1 < \Nfringe < 3$), which is located between the solid and dotted black lines.
As mentioned in the discussion of System 1 and as shown in the upper right panel of \cref{fig: compare LvsNP LvsRP td mcz}, precessional modulations are suppressed for face-on ($\hat{\mathbf{N}} \parallel \hat{\mathbf{J}}$) binary orientations \cite{TamanRP2025}.
This leads to a smaller improvement in match between lensed sources and RP templates as seen in the left panel of \cref{fig:compare_LvsRP_I_td_z1_mcz15_allsys}.  
Systems 2 and 3 do not exhibit the same suppression of precessional modulations as seen in the middle and right panels of Figs.~3 and 4 of \citeauthor{TamanRP2025}~\cite{TamanRP2025}, so RP templates can better mimic lensing signatures.
This yields the larger match improvements seen in the middle and right panels.
Precisely edge-on systems ($\hat{\mathbf{N}} \perp \hat{\mathbf{J}}$) like System 2 exhibit more regular precessional modulations than generically oriented systems like System 3 (compare the middle and right panels of Figs.~3 and 4 of \citeauthor{TamanRP2025}~\cite{TamanRP2025}).
This explains the more faithful correspondence between the predicted and observed locations of the peaks and troughs in the mismatch $\epsRP$ as a function of time delay $\tdel$ seen in the middle panel compared to the right panel.
Systems 2 and 3 nonetheless exhibit very similar mismatches as seen in \cref{fig: compare LvsNP LvsRP td mcz,fig:compare_LvsRP_I_td_z1_mcz15_allsys}, suggesting that conclusions drawn from the more easily interpreted System 2 are largely valid for generically oriented systems.

\subsection{Best-matching precession parameters}
\label{subsec: Best-matching RP}

\begin{figure*}[t]
    \centering
    \includegraphics[width=\textwidth]{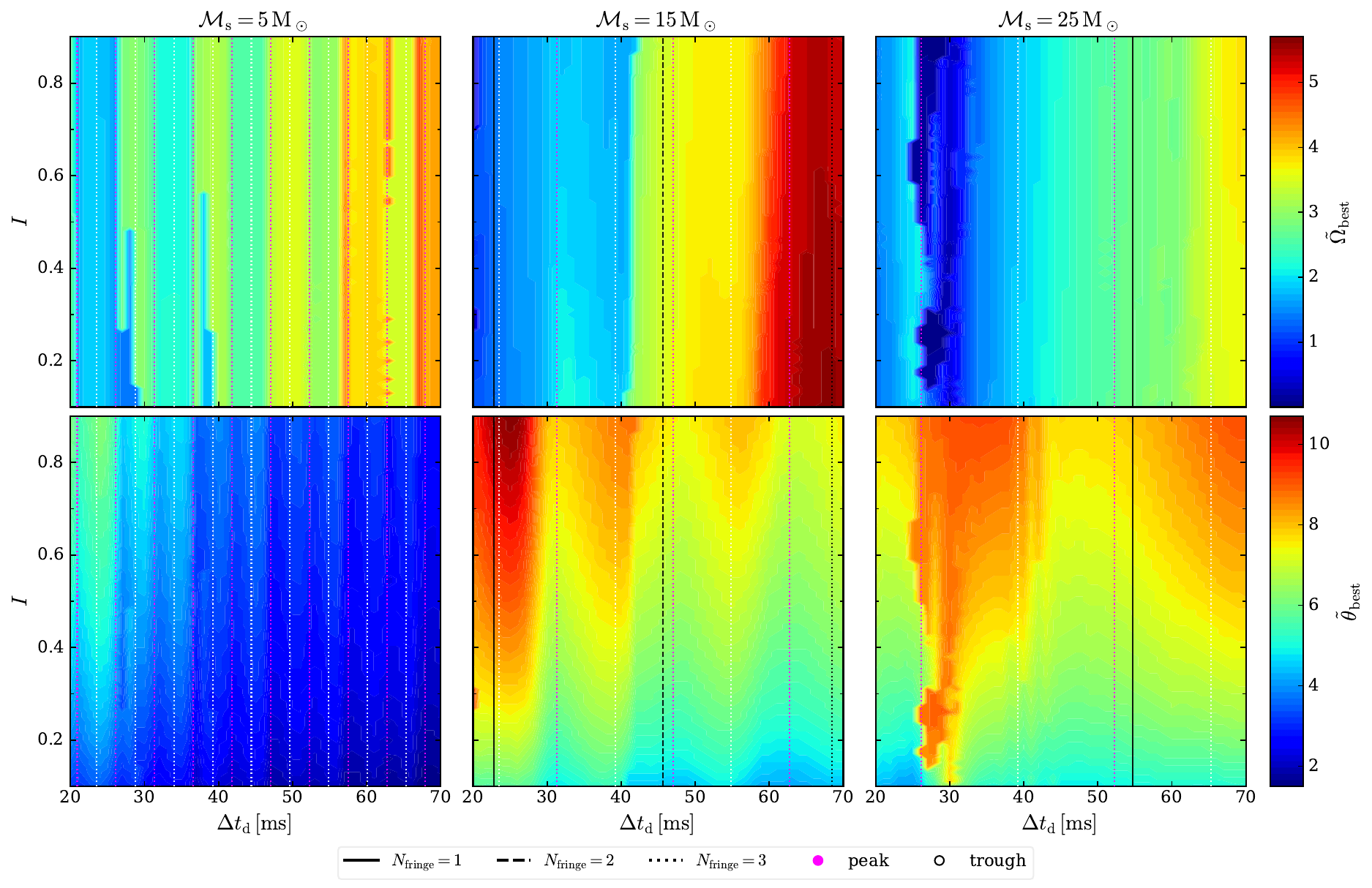}
    \caption{
    The best-fitting values of the precession frequency $\Omtilbest$ (top row) and amplitude $\thtilbest$ (bottom row) for RP templates as functions of the time delay $\tdel$ and flux ratio $I$ for two-image lensed sources.
    The left, middle, and right panels correspond to chirp masses of $\Mcs = 5$, $15$, and $25\,M_\odot$, and all sources are located at redshift $z = 1$ and have the binary orientation of System 2 listed in \cref{tab: source parameters_Lensed}.
    The mismatches $\epsRP$ associated with these parameter choices are shown in the bottom row of \cref{fig:compare_LvsNP_LvsRP_I_td_z1_edgeon}; as in that figure, the solid, dashed, and dotted black lines indicate $\Nfringe = 1$, 2, and 3 interference fringes in band, while the dotted magenta (white) lines mark the locations of peaks (troughs) in the mismatch between lensed sources and NP templates predicted by \cref{eq: mismatch extrema mcz}.
    }
    \label{fig:bestfit_prec_params_z1_edgeon}
\end{figure*}

\Cref{fig:bestfit_prec_params_z1_edgeon} shows the values of the precession frequency $\Omtilbest$ (top row) and amplitude $\thtilbest$ (bottom row) of the RP templates that best fit two-image lensed sources as functions of the lensing time delays $\tdel$ and flux ratios $I$.
The three columns have chirp masses $\Mcs = 5$, $15$, and $25\,M_\odot$ that for $20\,\mathrm{ms} < \tdel < 70\,\mathrm{ms}$ primarily correspond to Regions III, II, and I discussed in \cref{subsec: NP vs RP} above.
Two general trends are apparent for all three chirp masses.
In the top row, $\Omtilbest$ increases with the time delay $\tdel$, i.e., the contours go from blue to red as one moves from left to right.
Increasing $\tdel$ increases the number $\Nfringe$ of interference fringes in band according to \cref{eq: number of lensing modulations}.
The precession frequency $\Omtilbest$ of the best-fitting RP template must correspondingly increase to produce enough precessional modulations to mimic this higher number of fringes.
In the bottom row, $\thtilbest$ increases with the flux ratio $I$, i.e., the contours go from blue to red as one moves from bottom to top.
Increasing $I$ increases the amplitude of the lensing-induced interference fringes, and the precession amplitude $\thtilbest$ of the best-fitting RP template must increase to match them.

Both of these trends are clearest in the middle panels with $\Mcs = 15\,M_\odot$, because these panels are dominated by Region II in which there is a well-defined minimum of the mismatch $\epsRP$ as a function of $\Omtil$ and $\thtil$, as seen in the middle left panel of \cref{fig:sys2_combined_contour_waveform}.
The left panels, with $\Mcs = 5\,M_\odot$, are entirely in Region III for the range of time delays $\tdel$ shown.
These long lensed waveforms are poorly fit by RP templates because of their secular increase in the GW phase which scales quadratically with the precession amplitude $\thtil$, as discussed in \cref{appendix sec: secular phase}.
This accounts for the low values of $\thtilbest$ in the bottom left panel except for the lowest values of $\tdel$ which approach the boundary of Region II ($\Nfringe \approx 3$) where the positive correlation between $I$ and $\thtilbest$ is restored.

The right panels, with $\Mcs = 25\,M_\odot$, exhibit a strange feature near $\tdel \approx 30\,\mathrm{ms}$ that bucks the previously noted trends.
This feature is located in Region I ($\Nfringe < 1$), where the lensing-precession degeneracy is strongest and broad minima in the mismatch $\epsRP$ as a function of $\Omtil$ and $\thtil$ exist, as seen in the bottom left panel of \cref{fig:sys2_combined_contour_waveform}.
The feature results from a sharp transition in the global minimum between two widely separated but nearly degenerate local minima.
The values of $\Omtilbest$ and $\thtilbest$ are anti-correlated on both sides of the feature; as the precession frequency decreases from $\Omtilbest \approx 2$ to $\approx 1$ across the left boundary of the feature, the precession amplitude increases from $\thtilbest \approx 6$ to $\approx 8$.
This is roughly consistent with conserving the precessional contribution to the secular GW phase across the transition, which is proportional to $\Omtil\thtil^2$ as discussed in \cref{appendix sec: secular phase}.
With the exception of this feature, the right panels also exhibit the positive correlations between $\tdel$ and $\Omtilbest$ and between $I$ and $\thtilbest$ seen in the middle panels, particularly to the right of the solid black line at $\tdel \approx 55\,\mathrm{ms}$ marking the boundary of Region II.

\begin{figure*}[t]
    \centering
    \includegraphics[width=\textwidth]{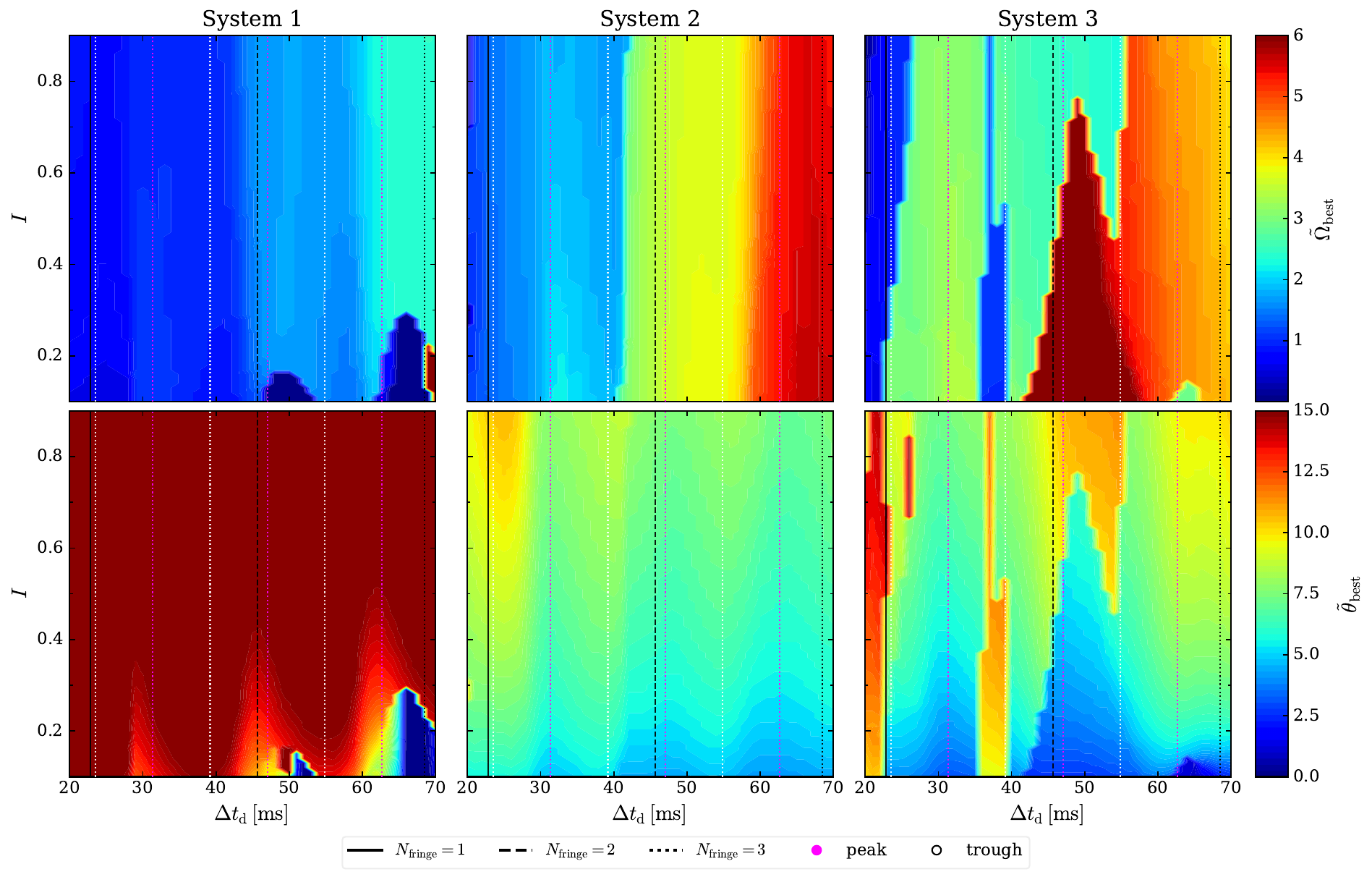}
    \caption{
    The best-fitting values of the precession frequency $\Omtilbest$ (top row) and amplitude $\thtilbest$ (bottom row) for RP templates as functions of the time delay $\tdel$ and flux ratio $I$ for two-image lensed sources.
    The left, middle, and right panels correspond to the binary orientations of Systems 1, 2, and 3 listed in \cref{tab: source parameters_Lensed}, and all sources have a chirp mass of $\Mcs = 15\,M_\odot$ and are located at redshift $z = 1$.
    The mismatches $\epsRP$ associated with these parameter choices are shown in \cref{fig:compare_LvsRP_I_td_z1_mcz15_allsys}; as in that figure, the solid, dashed, and dotted black lines indicate $\Nfringe = 1$, 2, and 3 interference fringes in band, while the dotted magenta (white) lines mark the locations of peaks (troughs) in the mismatch between lensed sources and NP templates predicted by \cref{eq: mismatch extrema mcz}.}
    \label{fig:bestfit_prec_params_z1_mcz15_allsys}
\end{figure*}

\Cref{fig:bestfit_prec_params_z1_mcz15_allsys} shows the values of the precession frequency (top row) and amplitude (bottom row) for the best-fitting RP templates whose mismatches with lensed sources in the binary orientations of Systems 1, 2, and 3 listed in \cref{tab: source parameters_Lensed} were shown in \cref{fig:compare_LvsRP_I_td_z1_mcz15_allsys}.
The middle panels corresponding to the edge-on ($\hat{\mathbf{N}} \perp \hat{\mathbf{J}}$) System 2 are the same as those in \cref{fig:bestfit_prec_params_z1_edgeon}, although the color bars have been changed to allow for a wider range of parameter values.
The left panels show $\Omtilbest$ and $\thtilbest$ for the face-on ($\hat{\mathbf{N}} \parallel \hat{\mathbf{J}}$) System 1.
For such face-on systems, RP templates struggle to match lensed source waveforms because of the suppressed precessional modulation of the GW amplitude and phase.
We see that for most time delays $\tdel$ and flux ratios $I$, the precession amplitude is driven to artificially high values ($\thtilbest \approx 15$, i.e., $\langle \thLJ \rangle \approx 1.5$~rad at $f = \fcut$) in a largely unsuccessful effort to match the lensing-induced interference fringes.
For the largest time delays and the smallest flux ratios in the bottom right corner, RP templates become completely ineffective and $\Omtilbest,\,\thtilbest \to 0$ corresponding to an NP template.

The right panels of \cref{fig:bestfit_prec_params_z1_mcz15_allsys} show the $\Omtilbest$ and $\thtilbest$ for the generically oriented System 3.
Although the minimum mismatch $\epsRP$ shown in the right panel of \cref{fig:compare_LvsRP_I_td_z1_mcz15_allsys} is a continuous function of the time delays $\tdel$ and flux ratios $I$ of the lensed sources, $\Omtilbest$ and $\thtilbest$ exhibit several discontinuous features reminiscent of the one seen in the right panels of \cref{fig:bestfit_prec_params_z1_edgeon}.
These occur even though the chirp mass and range of time delays place most of these panels in Region II that has well-defined minima of the mismatch $\epsP$ of \cref{E:epsP} with respect to $\Omtil$ and $\thtil$ for System 2.
The explanation is that the precessional modulations are less regular for the generically oriented System 3 than the precisely edge-on System 2, as can be seen by comparing the middle and right panels of Figs.~3 and 4 of \citeauthor{TamanRP2025}~\cite{TamanRP2025}.
This irregularity implies that when minimizing $\epsilon(\Omtil,\thtil,\gammaP)$ with respect to the nuisance phase $\gammaP$, numerous local minima of $\epsP$ are possible even when the number of interference fringes satisfies the criterion $1 < \Nfringe < 3$ that defines Region II.
These local minima typically have very similar secular contributions to the GW phase, which is proportional to $\Omtil\thtil^2$, as discussed in \cref{appendix sec: secular phase}.
Discontinuous transitions of the global minimum between these local minima as the time delay and flux ratio vary account for the discontinuous features in the right panels of \cref{fig:bestfit_prec_params_z1_mcz15_allsys}.
These discontinuities do not propagate to the mismatch itself for System 3 as can be seen in the bottom right panel of \cref{fig: compare LvsNP LvsRP td mcz} and the right panel of \cref{fig:compare_LvsRP_I_td_z1_mcz15_allsys}, supporting our conclusion that a strong lensing-precession degeneracy persists in Region II even for generically oriented systems.

\section{Discussion} \label{sec: Discussion}

\citeauthor{2025_Goyal_GW231123}~\cite{2025_Goyal_GW231123} recently suggested that the event GW231123 \cite{2025ApJ_GW23} could be interpreted as a lower-mass event that has been magnified by a foreground gravitational lens.
As the sensitivity of GW detectors increases in the future with the $A^\sharp$ upgrade~ \cite{Asharp} of current LVK facilities and third-generation GW detectors like the Einstein Telescope \cite{Maggiore2020ET} and Cosmic Explorer \cite{Evans2021CE}, we will observe GW signals at greater luminosity distances.
This will increase the probability that a fraction of these signals will be gravitationally lensed.
Strong lensing can produce multiple images that can interfere with each other if observed in band simultaneously.
Alternating constructive and destructive interference creates interference fringes in the observed signal.
The magnitude and separation of these fringes are determined by the time delay and flux ratio of the images, which themselves are set by the lensing potential and geometry.
There will be a non-zero mismatch between lensed waveforms possessing interference fringes and unmodulated, unlensed template waveforms, allowing lensing to be identified in GW signals.

The LVK collaboration has observed close to 400 GW candidates \cite{GWTC5}, the vast majority of which consist of BBHs.
In a generic BBH system, each black hole has a spin that need not be aligned with the orbital angular momentum of the binary.
The BBH population observed by the LVK collaboration has a distribution of effective precession spins $\chi_p$ \cite{chiP} that "peaks at a small but non-zero value of $\chi_p$ indicating that a non-negligible fraction of the BBH population have in-plane spins and experience precession"~\cite{GWTC5prop}.
This precession of the orbital angular momentum about the total angular momentum modulates the amplitude and phase of GWs emitted by such systems~\cite{Apostolatos1994}.
In order to claim to detect lensing in a GW event, the signature of lensing-induced interference fringes must be robustly distinguished from the more generic precessional modulation expected in BBH systems.

Investigating the distinguishability of lensing and precession is the primary goal of this paper.
To do so, we introduce a simple model of an unlensed, non-precessing inspiral waveform in \cref{subsec: GW background_LVP}, then describe how this waveform is modulated by gravitational lensing and precession in \cref{subsec: lensing background} and \cref{subsec: Precession background}, respectively.
The two-image amplification factors we consider are parameterized by the time delay $\tdel$ and flux ratio $I$ between the images, while the regularly precessing waveforms are parameterized by their dimensionless precession amplitude $\thtil$ and frequency $\Omtil$ (and a nuisance initial precession phase $\gammaP$)~\cite{GangardtSteinle2021,TamanRP2025}.
We calculate the mismatch between lensed source waveforms and NP/RP templates as described in \cref{subsec: match-filtering}.
From these mismatches, the Lindblom criterion of \cref{eq: Lindblom_LVP} provides an estimate of the source SNR needed to identify lensing; this SNR increases once precession is considered to the extent that RP templates have lower mismatches with lensed sources than NP templates.

We compare the mismatches of NP and RP templates in detail in \cref{sec: Results}.
We find that the degeneracy between lensed and precessing waveforms is primarily determined by the number of interference fringes $\Nfringe$ appearing in band as seen in \cref{fig: compare LvsNP LvsRP td mcz}.
If only a fraction of a fringe appears in band ($\Nfringe < 1$), there will be strong lensing-precession degeneracy quantified by a minimum mismatch between lensed sources and RP templates that is a factor $\mathcal{O}(100)$ lower than the mismatch between NP templates and the same lensed sources, as seen in the bottom panels of \cref{fig:sys2_combined_contour_waveform}.
If a few fringes appear in band ($1 < \Nfringe < 3$), the mismatch with RP templates will be reduced by a factor of order unity compared to NP templates, as in the middle panels of \cref{fig:sys2_combined_contour_waveform}.
Long waveforms with many interference fringes in band ($\Nfringe > 3$) will have minimal lensing-precession degeneracy, as in the top panels of \cref{fig:sys2_combined_contour_waveform}, because precessional modulations are spaced differently than interference fringes, and multiple precessional oscillations are typically accompanied by a large secular GW phase accumulation that matches poorly with lensed waveforms.
Nearly face-on ($\hat{\Vec{J}} \parallel \hat{\Vec{N}}$) systems are an exception to these conclusions, as precession yields negligible oscillations for such systems for all but the highest precession amplitudes.

This study reveals that precessional modulation presents a significant but not insurmountable obstacle to identifying gravitationally lensed BBH systems.
Further efforts to address this obstacle are certainly warranted, as the discovery of such lensed BBH systems would have considerable scientific payoffs.
Failure to identify lensing would bias parameter estimation for the sources, as~\citeauthor{2025_Goyal_GW231123}~\cite{2025_Goyal_GW231123} has claimed has occurred for the event GW231123, and as discussed generically in the statistical study of~\citeauthor{Oguri2018}~\cite{Oguri2018}.
Perhaps even more importantly, GW events are uniquely sensitive to strong lensing with the tiny time delays $\tdel < 0.1\,\mathrm{s}$ considered in this paper.
Such time delays would typically result from compact lenses with masses $M_L \lesssim 10^4\,M_\odot$ such as intermediate-mass black holes or primordial dark-matter halos.
Since competing dark-matter models predict markedly different populations of low-mass halos and subhalos, as outlined by~\citeauthor{BullockBoylanKolchin2017}~\cite{BullockBoylanKolchin2017}, identifying or constraining these structures offers a powerful probe of the nature of dark matter.
The hunt for such exotic objects provides yet another scientific motivation (as if more were needed) for continuing and expanding GW searches.

\acknowledgments
T.~N.~N.-V. was a post-baccalaureate fellow in the 2023--2024 TEXAS Bridge Program at the University of Texas at Dallas, which was funded by the NSF Partnerships in Astronomy and Astrophysics for Research and Education grant AST-2219128.
T.~S. was supported by NSF Gravitational Physics grant PHY-2309320, and B.~M. was supported by NSF Research Experiences for Undergraduates grant PHY-2348872.
M.~K. and L.~K. were supported by all three of these grants and gratefully acknowledge the confidence shown by the NSF in their efforts in research and education.

\appendix

\section{Oscillations in the mismatch between lensed and unlensed NP waveforms as a function of time delay and chirp mass}

\label{appendix sec: analytical mismatch}

The top left panel of \cref{fig: compare LvsNP LvsRP td mcz} reveals distinctive oscillations in the mismatch between lensed and unlensed NP waveforms as a function of the time delay $\tdel$ of the lensed waveform and their shared chirp mass $\Mcs$.
These oscillations can be understood by examining the mismatch $\epsilon = 1 - \mathrm{M}$, where
\begin{align}
\mathrm{M} &= \frac{\underset{\Delta t_c, \Delta\phi_c}{\max} \left\{ \langle \hL | \hUL \rangle \right\}}{\langle \hL | \hL \rangle^{1/2} \langle \hUL | \hUL \rangle^{1/2}} = \frac{\underset{\Delta t_c, \Delta\phi_c}{\max} \left\{  \mathcal{A} \right\}}{(\mathcal{BC})^{1/2}},
\end{align}
with
\begin{align}
\mathcal{A} &\equiv \int_{x_{\mathrm{min}}}^{x_{\mathrm{cut}}} dx\,\frac{x^{-7/3}[\cos{\Delta \Psi} + I^{1/2} \sin{( x + \Delta \Psi)}]}{S_n(x/2\pi\tdel)} , \\
\mathcal{B} &\equiv \int_{x_{\mathrm{min}}}^{x_{\mathrm{cut}}} dx\,\frac{x^{-7/3}(1 + I + 2I^{1/2} \sin x)}{S_n(x/2\pi\tdel)}, \\
\mathcal{C} &\equiv \int_{x_{\mathrm{min}}}^{x_{\mathrm{cut}}} dx\,\frac{x^{-7/3}}{S_n(x/2\pi\tdel)},
\end{align}
$x \equiv 2\pi f\tdel$, and $\Delta\Psi \equiv 2\pi f \Delta t_c - \Delta\phi_c$.
The integrals $\mathcal{B}$ and $\mathcal{C}$ are positive definite, but the integral $\mathcal{A}$ is not.
For the simple two-image lenses considered in this paper, $I < 1$, so $\mathcal{A}$ is maximized for $\Delta t_c,\,\Delta\phi_c \to 0$ for $\Nfringe \gtrsim 1$.
We adopt the analytic aLIGO power spectral density~\cite{Satyaprakash2009},
\begin{equation}
    \frac{S_n(y)}{S_0} = y^{-4.14} - 5y^{-2} + \frac{111(1 - y^{2} + 0.5y^{4})}{1 + 0.5y^{2}},
\end{equation}
where $y \equiv f/f_0$, $f_0 = 215\,\mathrm{Hz}$, $S_0 = 10^{-49}\,\mathrm{Hz}^{-1}$, and the lower frequency cutoff is $\fmin = 20\,\mathrm{Hz}$.
For the massive BBHs considered in this paper, $S_n$ is a monotonically decreasing function of frequency for $f < \fcut$.
This implies that the integral $\mathcal{A}$ is dominated by its upper limit and maximized when $\sin x_{\mathrm{cut}}$ passes through zero from above, i.e., for $x_{\mathrm{cut}} = (2n+1)\pi$ for integer $n$.
From the definitions of $x$ and $\fcut$ given by \cref{eq: fcut_LVP}, this implies that the peaks and troughs of the mismatch $\epsilon = 1 - \mathrm{M}$ will occur at 
\begin{equation} \label{eq: mismatch extrema mcz}
    \Mcs = 
        \begin{cases}
            \displaystyle \frac{\eta^{3/5} \tdel}{6^{3/2} \pi(1+z)n} & \text{for peaks,} \\[1em]
            \displaystyle \frac{\eta^{3/5} \tdel}{6^{3/2} \pi(1+z)(n + 1/2)} & \text{for troughs.}
    \end{cases}
\end{equation}
Comparing the magenta and white dotted lines with the colored contours in the top left panel of \cref{fig: compare LvsNP LvsRP td mcz} shows that this crude estimate is highly accurate for $n \geq 2$.

\begin{figure}
    \centering
    \includegraphics[width=1\columnwidth]{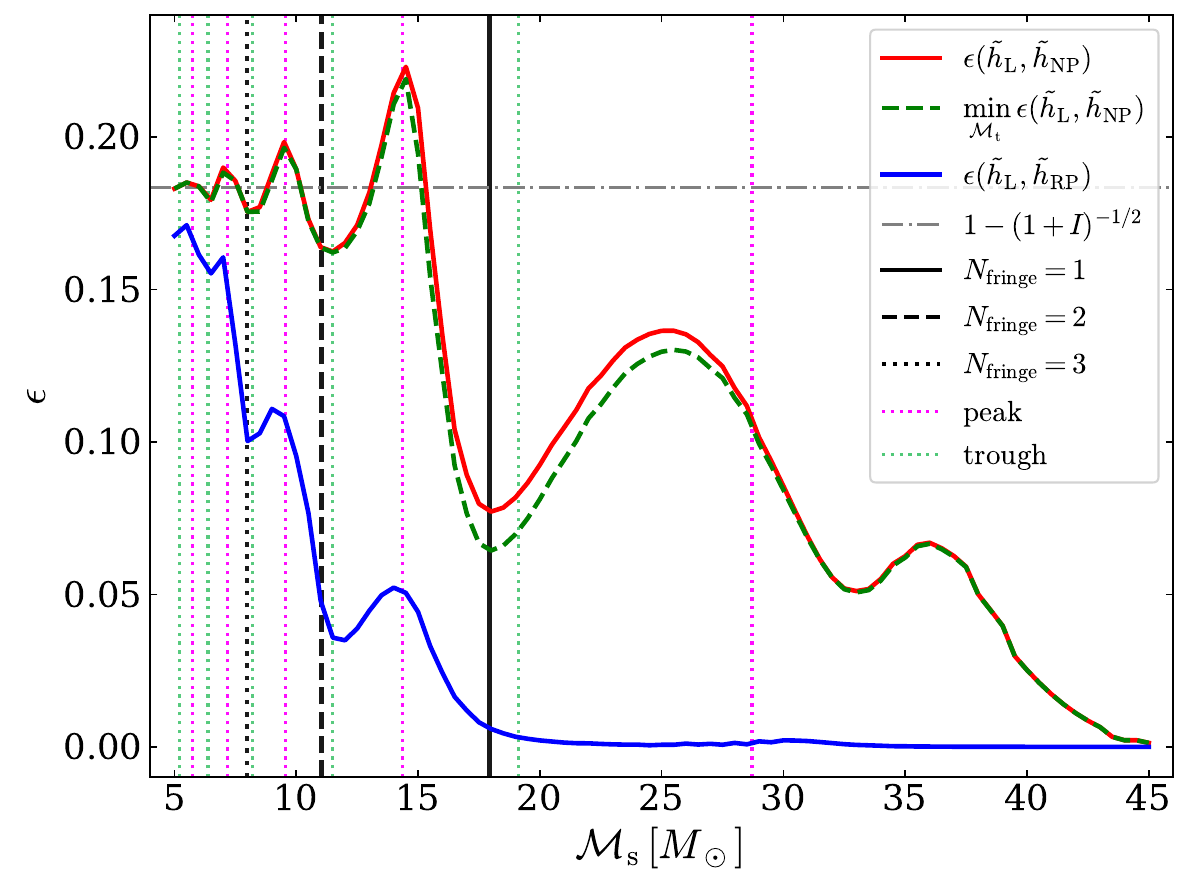}
    \caption{
    Mismatch $\epsilon$ as a function of source chirp mass $\Mcs$ for System~2 lensed binaries at fixed flux ratio $I=0.5$, time delay $\tdel = 30\,\mathrm{ms}$, and source redshift $z=1$.
    The solid red curve shows $\epsilon$ between a lensed NP source and an unlensed NP template with the same chirp mass, while the dashed green curve shows $\epsilon$ minimized with respect to the template chirp mass.
    The solid blue curve shows $\epsilon$ between a lensed NP source and an unlensed RP template minimized with respect to the precession parameters $\Omtil$, $\thtil$, and $\gammaP$.
    The solid, dashed, and dotted vertical black lines mark $\Nfringe=1$, $2$, and $3$ by \cref{eq: number of lensing modulations}.
    The dotted vertical magenta (lime-green) lines indicate predictions for the mismatch peaks (troughs) from \cref{eq: mismatch extrema mcz}.
    The dot-dashed gray horizontal line marks the asymptotic limit $1-(1+I)^{-1/2}$ as $\tdel\to\infty$.
    }
    \label{fig:sys2_mismatch_mcz}
\end{figure}

In \cref{fig:sys2_mismatch_mcz}, we investigate the predictions of \cref{eq: mismatch extrema mcz} for a typical time delay $\tdel = 30\,\mathrm{ms}$ and flux ratio $I = 0.5$.
We again see that these predictions are quite precise for $n \geq 2$, i.e., except for the very rightmost vertical dotted magenta and lime-green lines.
This figure also shows very strong lensing-precession degeneracy (blue curve approaches zero) in Region I (right of solid black line), modest degeneracy (blue curve a factor of few below red curve) in Region II (between dotted and solid black lines), and weak degeneracy (blue curve approaches red curve) in Region III (left of the dotted black line).
In the limit $\Mcs \to 0$, $x_{\mathrm{cut}} \to \infty$ and the integrals of the terms involving sinusoids of $x$ vanish in integrals $\mathcal{A}$ and $\mathcal{B}$.
This implies $\mathcal{A} = \mathcal{B}/(1 + I) = \mathcal{C}$ and thus $\mathrm{M} \to (1 + I)^{-1/2}$ and $\epsilon \to 1 - (1 + I)^{-1/2}$ (the red curve approaches the dot-dashed gray horizontal line) as $\Mcs \to 0$.
This limiting value for the mismatch was previously found in \citeauthor{Ali2023}~\cite{Ali2023}.

Finally, the close agreement between the solid red and dashed green curves shows that minimizing the mismatch with respect to the chirp mass of the template yields little reduction, indicating a lack of degeneracy between lensing and parameters that do not induce oscillatory features in the waveform.
This justifies our approximation of keeping such features constant between source and template waveforms in this paper.

\section{Dependence of secular GW phase accumulation on precession parameters}
\label{appendix sec: secular phase}

The two precessional terms in the GW phase given in \cref{eq: hf A and phase_LVP}, $\phi_p$ and $2\delta\Phi$, each depend on all three precession parameters $(\Omtil, \thtil, \gammaP)$ and provide both secular (monotonic) and oscillatory contributions.
To isolate the secular phase contribution to the GW phase, we study the phase accumulation for a binary with orientation and sky location specified as System 1 in~\cref{tab: source parameters_Lensed}.
As seen from Fig.~3 in \citeauthor{TamanRP2025}~\cite{TamanRP2025}, the oscillations in such face-on ($\hat{\Vec{J}} \parallel \hat{\Vec{N}}$) systems are suppressed, making it the ideal choice to study the secular phase contribution.

We begin with isolating the precessional phase contributions, which are given by the polarization phase $\phi_p$ in \cref{eq: phip_LVP} and the GW phase correction for precessing systems $\delta\Phi$ in \cref{eq: delta correction_LVP} (expanded in Eq.~(A19) in \citeauthor{TamanRP2025}~\cite{TamanRP2025}).
For System 1, the dot product in \cref{eq: LdotN_LVP} reduces to $\hat{\Vec{L}} \cdot \hat{\Vec{N}} = \cos\thLJ$, and the polarization angle $\psi$ from Eq.~(A14)~\cite{TamanRP2025} is $\psi = \PhiLJ + \Omega_{XH} = \PhiLJ$. 
Here $\Omega_{XH}$ is the longitude of ascending node (in the source frame from~\cite{TamanRP2025}) that is zero for System 1.

\Cref{eq: phip_LVP} can then be written as
\begin{align} 
\phi_p &= \tan^{-1}\left[ \frac{2\cos\thLJ \tan2\PhiLJ}{1 +\cos^2\thLJ} \right]\,,
\end{align}
and Eq.~(A19)~\cite{TamanRP2025} can be written as
\begin{align}
    \frac{d\delta\Phi}{df} &= - \left(\frac{\cos\thLJ}{\sin^2\thLJ}\right) \Bigg[ \langle\OmLJ\rangle \left( \frac{df}{dt} \right)^{-1} \sin^2\thLJ \Bigg] \,, \notag \\
    &=-\cos\thLJ\langle\OmLJ\rangle \left( \frac{df}{dt} \right)^{-1} \,.
\end{align} 

The total precessional contribution to the GW phase is then given by
\begin{align}
    \phi_p + 2\delta\Phi &= \tan^{-1}\left[ \frac{2\cos\thLJ \tan2\PhiLJ}{1 +\cos^2\thLJ} \right] \notag \\
    & \qquad - 2 \int_{\fmin}^f \cos\thLJ\langle\OmLJ\rangle \left( \frac{df'}{dt} \right)^{-1} df' \,. \label{eq: secular phase}
\end{align}
Expanding this result to quadratic order in $\thLJ$ yields
\begin{align}
    \phi_p + 2\delta\phi &\approx 2\PhiLJ - 2 \int_{\fmin}^f \left(1 - \frac{\thLJ^2}{2} \right)\langle\OmLJ\rangle \left( \frac{df'}{dt} \right)^{-1} df' \,, \notag \\
    &\approx 2\gammaP + \int_{\fmin}^f \thLJ^2 \langle\OmLJ\rangle \left( \frac{df'}{dt} \right)^{-1} df'\,.
\end{align}
So, the total secular phase accumulated over the binary inspiral in this limit is,
\begin{align}
    \Delta(\phi_p + 2\delta\Phi) &\equiv (\phi_p + 2\delta\Phi)|_{\fmin}^{\fcut} \notag \,, \\ 
    &\approx \int_{\fmin}^{\fcut} \thLJ^2 \langle\OmLJ\rangle \left( \frac{df'}{dt} \right)^{-1} df'\,, \notag \\
    &\approx C\Omtil\thtil^2 \,.\label{eq: secular phase fit2}
\end{align}
where $C$ is a constant.

\begin{figure*}[t]
    \centering
    \includegraphics[width=\textwidth]{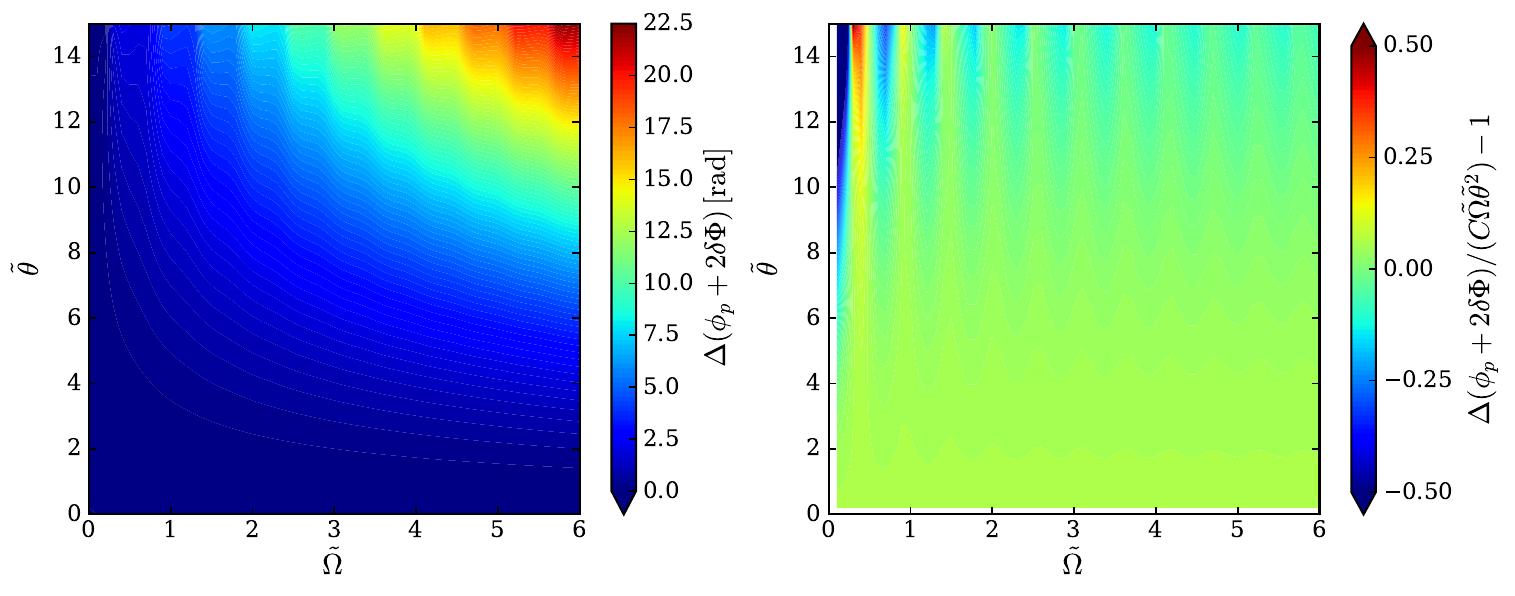}
    \caption{\textbf{Left}: Precessional contribution $\Delta(\phi_p + 2\delta\Phi)$ to the GW phase accumulated during the inspiral given by \cref{eq: secular phase} as a function of the dimensionless precession frequency $\Omtil$ and amplitude $\thtil$ for BBHs with source chirp mass $\Mcs = 15\,M_\odot$ at redshift $z=1$ with the orientation of System 1 listed in \cref{tab: source parameters_Lensed}.
    \textbf{Right}: Fractional residual GW phase after the numerical fit $\Delta(\phi_p + 2\delta\Phi) \approx C\Omtil\thtil^2$ with $C = 1.78 \times 10^{-2}$ has been subtracted from the precessional contribution $\Delta(\phi_p + 2\delta\Phi)$ shown in the left panel.}
    \label{fig: secular phase RP}
\end{figure*}

The left panel of \cref{fig: secular phase RP} shows the GW phase accumulated according to \cref{eq: secular phase} during the inspiral of BBHs with source chirp mass of $\Mcs = 15\,M_\odot$ at redshift $z=1$ with the orientation of System 1.
This phase accumulation is almost entirely secular for small precession amplitudes, but the oscillatory contributions grow as $\thtil$ increases.
If we numerically fit the approximation $\Delta(\phi_p + 2\delta\Phi) \approx C\Omtil\thtil^2$ to these results, we find a best-fit value of $C = 1.78 \times 10^{-2}$. 
The right panel in \cref{fig: secular phase RP} shows the fractional residual after subtracting this approximation from the exact result given by \cref{eq: secular phase}.
This residual is small except for extremely high precession amplitudes $\thtil \gtrsim 7$, close to the 95\textsuperscript{th}-percentile value $\thtil \approx 8.05$ for populations of equal-mass binaries with maximal spins and isotropic orientations~\cite{TamanRP2025}.
The approximation is therefore valid over most of the physically realizable range of precession parameters $(\Omtil, \thtil, \gammaP)$, though not for the largest amplitudes explored in \cref{sec: Results}.

\newpage
\bibliography{main}

@ARTICLE{chiP,
       author = {{Hannam}, Mark and {Schmidt}, Patricia and {Boh{\'e}}, Alejandro and {Haegel}, Le{\"\i}la and {Husa}, Sascha and {Ohme}, Frank and {Pratten}, Geraint and {P{\"u}rrer}, Michael},
        title = "{Simple Model of Complete Precessing Black-Hole-Binary Gravitational Waveforms}",
      journal = {\prl},
         year = 2014,
        month = oct,
       volume = {113},
       number = {15},
          eid = {151101},
        pages = {151101},
          doi = {10.1103/PhysRevLett.113.151101},
archivePrefix = {arXiv},
       eprint = {1308.3271},
 primaryClass = {gr-qc},
       adsurl = {https://ui.adsabs.harvard.edu/abs/2014PhRvL.113o1101H}
}

@ARTICLE{GWTC5prop,
       author = {{ A.~G. {Abac} \it{et al.} }{(LIGO Scientific, Virgo, and KAGRA Collaborations)}},
        title = "{GWTC-5.0: Population Properties of Merging Compact Binaries}",
      journal = {arXiv e-prints},
         year = 2026,
        month = may,
          eid = {arXiv:2605.27226},
        pages = {arXiv:2605.27226},
          doi = {10.48550/arXiv.2605.27226},
archivePrefix = {arXiv},
       eprint = {2605.27226},
 primaryClass = {astro-ph.HE},
       adsurl = {https://ui.adsabs.harvard.edu/abs/2026arXiv260527226T}
}

@ARTICLE{Asharp,
       author = {{Gupta}, Ish and {Afle}, Chaitanya and {Arun}, K.~G. and {Bandopadhyay}, Ananya and {Baryakhtar}, Masha and {Biscoveanu}, Sylvia and {Borhanian}, Ssohrab and {Broekgaarden}, Floor and {Corsi}, Alessandra and {Dhani}, Arnab and {Evans}, Matthew and {Hall}, Evan D. and {Hannuksela}, Otto A. and {Kacanja}, Keisi and {Kashyap}, Rahul and {Khadkikar}, Sanika and {Kuns}, Kevin and {Li}, Tjonnie G.~F. and {Miller}, Andrew L. and {Harvey Nitz}, Alexander and {Owen}, Benjamin J. and {Palomba}, Cristiano and {Pearce}, Anthony and {Phurailatpam}, Hemantakumar and {Rajbhandari}, Binod and {Read}, Jocelyn and {Romano}, Joseph D. and {Sathyaprakash}, Bangalore S. and {Shoemaker}, David H. and {Singh}, Divya and {Vitale}, Salvatore and {Barsotti}, Lisa and {Berti}, Emanuele and {Cahillane}, Craig and {Chen}, Hsin-Yu and {Fritschel}, Peter and {Haster}, Carl-Johan and {Landry}, Philippe and {Lovelace}, Geoffrey and {McClelland}, David and {J J Slagmolen}, Bram and {R Smith}, Joshua and {Soares-Santos}, Marcelle and {Sun}, Ling and {Tanner}, David and {Yamamoto}, Hiro and {Zucker}, Michael},
        title = "{Characterizing gravitational wave detector networks: from A to cosmic explorer}",
      journal = {Classical and Quantum Gravity},
         year = 2024,
        month = dec,
       volume = {41},
       number = {24},
          eid = {245001},
        pages = {245001},
          doi = {10.1088/1361-6382/ad7b99},
archivePrefix = {arXiv},
       eprint = {2307.10421},
 primaryClass = {gr-qc},
       adsurl = {https://ui.adsabs.harvard.edu/abs/2024CQGra..41x5001G}
}

@ARTICLE{GWTC5,
       author = {{The LIGO Scientific Collaboration} and {the Virgo Collaboration} and {the KAGRA Collaboration}},
        title = "{GWTC-5.0: Observations from the Second Part of the Fourth LIGO-Virgo-KAGRA Observing Run and Updates to the Gravitational-Wave Transient Catalog}",
      journal = {arXiv e-prints},
         year = 2026,
        month = may,
          eid = {arXiv:2605.27225},
        pages = {arXiv:2605.27225},
          doi = {10.48550/arXiv.2605.27225},
archivePrefix = {arXiv},
       eprint = {2605.27225},
 primaryClass = {gr-qc},
       adsurl = {https://ui.adsabs.harvard.edu/abs/2026arXiv260527225T}
}

@ARTICLE{Shan2026,
       author = {{Shan}, Xikai and {Yang}, Huan and {Mao}, Shude and {Hannuksela}, Otto A.},
        title = "{Spin Precession Signatures as an Indicator of Microlensing in Strongly Lensed Gravitational Waves}",
      journal = {\apj},
         year = 2026,
        month = jun,
       volume = {1004},
       number = {1},
          eid = {35},
        pages = {35},
          doi = {10.3847/1538-4357/ae6ce1},
archivePrefix = {arXiv},
       eprint = {2508.21262},
 primaryClass = {astro-ph.CO},
       adsurl = {https://ui.adsabs.harvard.edu/abs/2026ApJ..1004...35S}
}

@ARTICLE{LiuKim2024,
       author = {{Liu}, Anna and {Kim}, Kyungmin},
        title = "{Can we discern millilensed gravitational-wave signals from signals produced by precessing binary black holes with ground-based detectors?}",
      journal = {\prd},
         year = 2024,
        month = dec,
       volume = {110},
       number = {12},
          eid = {123008},
        pages = {123008},
          doi = {10.1103/PhysRevD.110.123008},
archivePrefix = {arXiv},
       eprint = {2301.07253},
 primaryClass = {gr-qc},
       adsurl = {https://ui.adsabs.harvard.edu/abs/2024PhRvD.110l3008L}
}

@ARTICLE{Zhao2017,
       author = {{Zhao}, Xinyu and {Kesden}, Michael and {Gerosa}, Davide},
        title = "{Nutational resonances, transitional precession, and precession-averaged evolution in binary black-hole systems}",
      journal = {\prd},
         year = 2017,
        month = jul,
       volume = {96},
       number = {2},
          eid = {024007},
        pages = {024007},
          doi = {10.1103/PhysRevD.96.024007},
archivePrefix = {arXiv},
       eprint = {1705.02369},
 primaryClass = {gr-qc},
       adsurl = {https://ui.adsabs.harvard.edu/abs/2017PhRvD..96b4007Z}
}

@ARTICLE{Ali2023,
       author = {{Ali}, Saif and {Stoikos}, Evangelos and {Meade}, Evan and {Kesden}, Michael and {King}, Lindsay},
        title = "{Detectability of strongly lensed gravitational waves using model-independent image parameters}",
      journal = {\prd},
         year = 2023,
        month = may,
       volume = {107},
       number = {10},
          eid = {103023},
        pages = {103023},
          doi = {10.1103/PhysRevD.107.103023},
archivePrefix = {arXiv},
       eprint = {2210.01873},
 primaryClass = {gr-qc},
       adsurl = {https://ui.adsabs.harvard.edu/abs/2023PhRvD.107j3023A}
}

@ARTICLE{GangardtSteinle2021,
       author = {{Gangardt}, Daria and {Steinle}, Nathan and {Kesden}, Michael and {Gerosa}, Davide and {Stoikos}, Evangelos},
        title = "{A taxonomy of black-hole binary spin precession and nutation}",
      journal = {\prd},
         year = 2021,
        month = jun,
       volume = {103},
       number = {12},
          eid = {124026},
        pages = {124026},
          doi = {10.1103/PhysRevD.103.124026},
archivePrefix = {arXiv},
       eprint = {2103.03894},
 primaryClass = {gr-qc},
       adsurl = {https://ui.adsabs.harvard.edu/abs/2021PhRvD.103l4026G}
}

@ARTICLE{MultiTimeScale2015,
       author = {{Gerosa}, Davide and {Kesden}, Michael and {Sperhake}, Ulrich and {Berti}, Emanuele and {O'Shaughnessy}, Richard},
        title = "{Multi-timescale analysis of phase transitions in precessing black-hole binaries}",
      journal = {\prd},
         year = 2015,
        month = sep,
       volume = {92},
       number = {6},
          eid = {064016},
        pages = {064016},
          doi = {10.1103/PhysRevD.92.064016},
archivePrefix = {arXiv},
       eprint = {1506.03492},
 primaryClass = {gr-qc},
       adsurl = {https://ui.adsabs.harvard.edu/abs/2015PhRvD..92f4016G}
}

@ARTICLE{Satyaprakash2009,
       author = {{Sathyaprakash}, B.~S. and {Schutz}, Bernard F.},
        title = "{Physics, Astrophysics and Cosmology with Gravitational Waves}",
      journal = {Living Reviews in Relativity},
        year = 2009,
        month = dec,
       volume = {12},
       number = {1},
          eid = {2},
        pages = {2},
          doi = {10.12942/lrr-2009-2},
archivePrefix = {arXiv},
       eprint = {0903.0338},
 primaryClass = {gr-qc},
       adsurl = {https://ui.adsabs.harvard.edu/abs/2009LRR....12....2S}
}

@article{CutlerFlanagan1994,
  title = {Gravitational waves from merging compact binaries: How accurately can one extract the binary's parameters from the inspiral waveform?},
  author = {Cutler, Curt and Flanagan, \'Eanna E.},
  journal = {Phys. Rev. D},
  volume = {49},
  issue = {6},
  pages = {2658--2697},
  numpages = {0},
  year = {1994},
  month = {Mar},
  publisher = {American Physical Society},
  doi = {10.1103/PhysRevD.49.2658},
  url = {https://link.aps.org/doi/10.1103/PhysRevD.49.2658}
}

@article{Apostolatos1994,
  title = {Spin-induced orbital precession and its modulation of the gravitational waveforms from merging binaries},
  author = {Apostolatos, Theocharis A. and Cutler, Curt and Sussman, Gerald J. and Thorne, Kip S.},
  journal = {Phys. Rev. D},
  volume = {49},
  issue = {12},
  pages = {6274--6297},
  numpages = {0},
  year = {1994},
  month = {Jun},
  publisher = {American Physical Society},
  doi = {10.1103/PhysRevD.49.6274},
  url = {https://link.aps.org/doi/10.1103/PhysRevD.49.6274}
}

@ARTICLE{TakahashiNakamura2003,
       author = {{Takahashi}, Ryuichi and {Nakamura}, Takashi},
        title = "{Wave Effects in the Gravitational Lensing of Gravitational Waves from Chirping Binaries}",
      journal = {\apj},
         year = 2003,
        month = oct,
       volume = {595},
       number = {2},
        pages = {1039-1051},
          doi = {10.1086/377430},
archivePrefix = {arXiv},
       eprint = {astro-ph/0305055},
 primaryClass = {astro-ph},
       adsurl = {https://ui.adsabs.harvard.edu/abs/2003ApJ...595.1039T}
}

@ARTICLE{TamanRP2025,
       author = {{Singh}, Tamanjyot and {Stoikos}, Evangelos and {Ali}, Saif and {Steinle}, Nathan and {Kesden}, Michael and {King}, Lindsay},
        title = "{Detecting regular precession using a new gravitational waveform model directly parameterized by both precession amplitude and frequency}",
      journal = {arXiv e-prints},
         year = 2025,
        month = sep,
          eid = {arXiv:2509.10628},
        pages = {arXiv:2509.10628},
archivePrefix = {arXiv},
       eprint = {2509.10628},
 primaryClass = {gr-qc},
       adsurl = {https://ui.adsabs.harvard.edu/abs/2025arXiv250910628S}
}

@article{PhysRevD.52.848,
  title = {Gravitational waves from inspiraling compact binaries: Parameter estimation using second-post-Newtonian waveforms},
  author = {Poisson, Eric and Will, Clifford M.},
  journal = {Phys. Rev. D},
  volume = {52},
  issue = {2},
  pages = {848--855},
  numpages = {0},
  year = {1995},
  month = {Jul},
  publisher = {American Physical Society},
  doi = {10.1103/PhysRevD.52.848},
  url = {https://link.aps.org/doi/10.1103/PhysRevD.52.848}
}

@ARTICLE{1992ApJ...400..175B,
       author = {{Bildsten}, Lars and {Cutler}, Curt},
        title = "{Tidal Interactions of Inspiraling Compact Binaries}",
      journal = {\apj},
         year = 1992,
        month = nov,
       volume = {400},
        pages = {175},
          doi = {10.1086/171983},
       adsurl = {https://ui.adsabs.harvard.edu/abs/1992ApJ...400..175B}
}

@article{Finn1992,
    author = {{Finn}, Lee S.},
    title = "{Detection, measurement, and gravitational radiation}",
    journal = {\prd},
         year = 1992,
        month = dec,
       volume = {46},
       number = {12},
        pages = {5236-5249},
          doi = {10.1103/PhysRevD.46.5236},
archivePrefix = {arXiv},
       eprint = {gr-qc/9209010},
 primaryClass = {gr-qc},
       adsurl = {https://ui.adsabs.harvard.edu/abs/1992PhRvD..46.5236F}
}

@article{Wiseman1992,
  title = {Coalescing binary systems of compact objects to ${(\mathrm{post})}^{5/2}$-Newtonian order. II. Higher-order wave forms and radiation recoil},
  author = {Wiseman, Alan G.},
  journal = {Phys. Rev. D},
  volume = {46},
  issue = {4},
  pages = {1517--1539},
  numpages = {0},
  year = {1992},
  month = {Aug},
  publisher = {American Physical Society},
  doi = {10.1103/PhysRevD.46.1517},
  url = {https://link.aps.org/doi/10.1103/PhysRevD.46.1517}
}

@article{Lindblom2008,
  title = {Model waveform accuracy standards for gravitational wave data analysis},
  author = {Lindblom, Lee and Owen, Benjamin J. and Brown, Duncan A.},
  journal = {Phys. Rev. D},
  volume = {78},
  issue = {12},
  pages = {124020},
  numpages = {12},
  year = {2008},
  month = {Dec},
  publisher = {American Physical Society},
  doi = {10.1103/PhysRevD.78.124020},
  url = {https://link.aps.org/doi/10.1103/PhysRevD.78.124020}
}

@ARTICLE{Einstein1936,
       author = {{Einstein}, Albert},
        title = "{Lens-Like Action of a Star by the Deviation of Light in the Gravitational Field}",
      journal = {Science},
         year = 1936,
        month = dec,
       volume = {84},
       number = {2188},
        pages = {506-507},
          doi = {10.1126/science.84.2188.506},
       adsurl = {https://ui.adsabs.harvard.edu/abs/1936Sci....84..506E}
}

@ARTICLE{Lawrence1971,
       author = {{Lawrence}, J.~K.},
        title = "{Focusing of gravitational radiation by interior gravitational fields.}",
      journal = {Nuovo Cimento B Serie},
         year = 1971,
        month = jan,
       volume = {6B},
        pages = {225-235},
          doi = {10.1007/BF02735388},
       adsurl = {https://ui.adsabs.harvard.edu/abs/1971NCimB...6..225L}
}

@article{Nakamura1998,
  title = {Gravitational Lensing of Gravitational Waves from Inspiraling Binaries by a Point Mass Lens},
  author = {Nakamura, Takahiro T.},
  journal = {Phys. Rev. Lett.},
  volume = {80},
  issue = {6},
  pages = {1138--1141},
  numpages = {0},
  year = {1998},
  month = {Feb},
  publisher = {American Physical Society},
  doi = {10.1103/PhysRevLett.80.1138},
  url = {https://link.aps.org/doi/10.1103/PhysRevLett.80.1138}
}

@ARTICLE{Nakamura1999,
       author = {{Nakamura}, T.~T. and {Deguchi}, S.},
        title = "{Wave Optics in Gravitational Lensing}",
      journal = {Progress of Theoretical Physics Supplement},
         year = 1999,
        month = jan,
       volume = {133},
        pages = {137-153},
          doi = {10.1143/PTPS.133.137},
       adsurl = {https://ui.adsabs.harvard.edu/abs/1999PThPS.133..137N}
}

@ARTICLE{Evans2021CE,
       author = {{Evans}, Matthew and {Adhikari}, Rana X and {Afle}, Chaitanya and {Ballmer}, Stefan W. and {Biscoveanu}, Sylvia and {Borhanian}, Ssohrab and {Brown}, Duncan A. and {Chen}, Yanbei and {Eisenstein}, Robert and {Gruson}, Alexandra and {Gupta}, Anuradha and {Hall}, Evan D. and {Huxford}, Rachael and {Kamai}, Brittany and {Kashyap}, Rahul and {Kissel}, Jeff S. and {Kuns}, Kevin and {Landry}, Philippe and {Lenon}, Amber and {Lovelace}, Geoffrey and {McCuller}, Lee and {Ng}, Ken K.~Y. and {Nitz}, Alexander H. and {Read}, Jocelyn and {Sathyaprakash}, B.~S. and {Shoemaker}, David H. and {Slagmolen}, Bram J.~J. and {Smith}, Joshua R. and {Srivastava}, Varun and {Sun}, Ling and {Vitale}, Salvatore and {Weiss}, Rainer},
        title = "{A Horizon Study for Cosmic Explorer: Science, Observatories, and Community}",
      journal = {arXiv e-prints},
         year = 2021,
        month = sep,
          eid = {arXiv:2109.09882},
        pages = {arXiv:2109.09882},
          doi = {10.48550/arXiv.2109.09882},
archivePrefix = {arXiv},
       eprint = {2109.09882},
 primaryClass = {astro-ph.IM},
       adsurl = {https://ui.adsabs.harvard.edu/abs/2021arXiv210909882E}
}

@ARTICLE{Maggiore2020ET,
       author = {{Maggiore}, Michele and {Van Den Broeck}, Chris and {Bartolo}, Nicola and {Belgacem}, Enis and {Bertacca}, Daniele and {Bizouard}, Marie Anne and {Branchesi}, Marica and {Clesse}, Sebastien and {Foffa}, Stefano and {Garc{\'\i}a-Bellido}, Juan and {Grimm}, Stefan and {Harms}, Jan and {Hinderer}, Tanja and {Matarrese}, Sabino and {Palomba}, Cristiano and {Peloso}, Marco and {Ricciardone}, Angelo and {Sakellariadou}, Mairi},
        title = "{Science case for the Einstein telescope}",
      journal = {\jcap},
         year = 2020,
        month = mar,
       volume = {2020},
       number = {3},
          eid = {050},
        pages = {050},
          doi = {10.1088/1475-7516/2020/03/050},
archivePrefix = {arXiv},
       eprint = {1912.02622},
 primaryClass = {astro-ph.CO},
       adsurl = {https://ui.adsabs.harvard.edu/abs/2020JCAP...03..050M}
}

@article{liao2017precision,
  title = {Precision cosmology from future lensed gravitational wave and electromagnetic signals},
  author = {Liao, Kai and Fan, Xi-Long and Ding, Xuheng and Biesiada, Marek and Zhu, Zong-Hong},
  year = {2017},
  journal = {Nature Communications},
  publisher = {Springer Science and Business Media LLC},
  volume = {8},
  number = {1},
  doi = {10.1038/s41467-017-01152-9},
  url = {https://doi.org/10.1038/s41467-017-01152-9}
}

@ARTICLE{1979Natur.279..381W,
       author = {{Walsh}, D. and {Carswell}, R.~F. and {Weymann}, R.~J.},
        title = "{0957+561 A, B: twin quasistellar objects or gravitational lens?}",
      journal = {\nat},
         year = 1979,
        month = may,
       volume = {279},
        pages = {381-384},
          doi = {10.1038/279381a0},
       adsurl = {https://ui.adsabs.harvard.edu/abs/1979Natur.279..381W}
}

@ARTICLE{2008ApJ...682..964B,
       author = {{Bolton}, Adam S. and {Burles}, Scott and {Koopmans}, L{\'e}on V.~E. and {Treu}, Tommaso and {Gavazzi}, Rapha{\"e}l and {Moustakas}, Leonidas A. and {Wayth}, Randall and {Schlegel}, David J.},
        title = "{The Sloan Lens ACS Survey. V. The Full ACS Strong-Lens Sample}",
      journal = {\apj},
         year = 2008,
        month = aug,
       volume = {682},
       number = {2},
        pages = {964-984},
          doi = {10.1086/589327},
archivePrefix = {arXiv},
       eprint = {0805.1931},
 primaryClass = {astro-ph},
       adsurl = {https://ui.adsabs.harvard.edu/abs/2008ApJ...682..964B}
}

@ARTICLE{2003MNRAS.341...13B,
       author = {{Browne}, I.~W.~A. and {Wilkinson}, P.~N. and {Jackson}, N.~J.~F. and {Myers}, S.~T. and {Fassnacht}, C.~D. and {Koopmans}, L.~V.~E. and {Marlow}, D.~R. and {Norbury}, M. and {Rusin}, D. and {Sykes}, C.~M. and {Biggs}, A.~D. and {Blandford}, R.~D. and {de Bruyn}, A.~G. and {Chae}, K.-H. and {Helbig}, P. and {King}, L.~J. and {McKean}, J.~P. and {Pearson}, T.~J. and {Phillips}, P.~M. and {Readhead}, A.~C.~S. and {Xanthopoulos}, E. and {York}, T.},
        title = "{The Cosmic Lens All-Sky Survey - II. Gravitational lens candidate selection and follow-up}",
      journal = {\mnras},
         year = 2003,
        month = may,
       volume = {341},
       number = {1},
        pages = {13-32},
          doi = {10.1046/j.1365-8711.2003.06257.x},
archivePrefix = {arXiv},
       eprint = {astro-ph/0211069},
 primaryClass = {astro-ph},
       adsurl = {https://ui.adsabs.harvard.edu/abs/2003MNRAS.341...13B}
}

@ARTICLE{2024SSRv..220...58V,
       author = {{Vegetti}, S. and {Birrer}, S. and {Despali}, G. and {Fassnacht}, C.~D. and {Gilman}, D. and {Hezaveh}, Y. and {Perreault Levasseur}, L. and {McKean}, J.~P. and {Powell}, D.~M. and {O'Riordan}, C.~M. and {Vernardos}, G.},
        title = "{Strong Gravitational Lensing as a Probe of Dark Matter}",
      journal = {\ssr},
         year = 2024,
        month = aug,
       volume = {220},
       number = {5},
          eid = {58},
        pages = {58},
          doi = {10.1007/s11214-024-01087-w},
archivePrefix = {arXiv},
       eprint = {2306.11781},
 primaryClass = {astro-ph.CO},
       adsurl = {https://ui.adsabs.harvard.edu/abs/2024SSRv..220...58V}
}

@ARTICLE{2025arXiv251216347T,
       author = {{ A.~G. {Abac} \it{et al.} }{(LIGO Scientific, Virgo, and KAGRA Collaborations)}},
        title = "{GWTC-4.0: Searches for Gravitational-Wave Lensing Signatures}",
      journal = {arXiv e-prints},
         year = 2025,
        month = dec,
          eid = {arXiv:2512.16347},
        pages = {arXiv:2512.16347},
          doi = {10.48550/arXiv.2512.16347},
archivePrefix = {arXiv},
       eprint = {2512.16347},
 primaryClass = {gr-qc},
       adsurl = {https://ui.adsabs.harvard.edu/abs/2025arXiv251216347T}
}

@ARTICLE{2018arXiv180707062H,
       author = {{Haris}, K. and {Mehta}, Ajit Kumar and {Kumar}, Sumit and {Venumadhav}, Tejaswi and {Ajith}, Parameswaran},
        title = "{Identifying strongly lensed gravitational wave signals from binary black hole mergers}",
      journal = {arXiv e-prints},
         year = 2018,
        month = jul,
          eid = {arXiv:1807.07062},
        pages = {arXiv:1807.07062},
          doi = {10.48550/arXiv.1807.07062},
archivePrefix = {arXiv},
       eprint = {1807.07062},
 primaryClass = {gr-qc},
       adsurl = {https://ui.adsabs.harvard.edu/abs/2018arXiv180707062H}
}

@ARTICLE{2025_Goyal_GW231123,
       author = {{Goyal}, Srashti and {Villarrubia-Rojo}, Hector and {Zumalacarregui}, Miguel},
        title = "{Across the Universe: GW231123 as a magnified and diffracted black hole merger}",
      journal = {arXiv e-prints},
         year = 2025,
        month = dec,
          eid = {arXiv:2512.17631},
        pages = {arXiv:2512.17631},
          doi = {10.48550/arXiv.2512.17631},
archivePrefix = {arXiv},
       eprint = {2512.17631},
 primaryClass = {astro-ph.GA},
       adsurl = {https://ui.adsabs.harvard.edu/abs/2025arXiv251217631G}
}

@ARTICLE{2025_Chan_GW231123,
       author = {{Chan}, Juno C.~L. and {Mar{\'\i}a Ezquiaga}, Jose and {Lo}, Rico K.~L. and {Bowman}, Joey and {Maga{\~n}a Zertuche}, Lorena and {Vujeva}, Luka},
        title = "{Discovering gravitational waveform distortions from lensing: a deep dive into GW231123}",
      journal = {arXiv e-prints},
         year = 2025,
        month = dec,
          eid = {arXiv:2512.16916},
        pages = {arXiv:2512.16916},
          doi = {10.48550/arXiv.2512.16916},
archivePrefix = {arXiv},
       eprint = {2512.16916},
 primaryClass = {gr-qc},
       adsurl = {https://ui.adsabs.harvard.edu/abs/2025arXiv251216916C}
}

@ARTICLE{2025ApJ_GW23,
       author = {{ A.~G. {Abac} \it{et al.} }{(LIGO Scientific, Virgo, and KAGRA Collaborations)}},
        title = "{GW231123: A Binary Black Hole Merger with Total Mass 190─265 M$_{{\ensuremath{\odot}}}$}",
      journal = {\apjl},
         year = 2025,
        month = nov,
       volume = {993},
       number = {1},
          eid = {L25},
        pages = {L25},
          doi = {10.3847/2041-8213/ae0c9c},
archivePrefix = {arXiv},
       eprint = {2507.08219},
 primaryClass = {astro-ph.HE},
       adsurl = {https://ui.adsabs.harvard.edu/abs/2025ApJ...993L..25A}
}

@ARTICLE{Cluster_rev_2024SSRv..220...19N,
       author = {{Natarajan}, P. and {Williams}, L.~L.~R. and {Brada{\v{c}}}, M. and {Grillo}, C. and {Ghosh}, A. and {Sharon}, K. and {Wagner}, J.},
        title = "{Strong Lensing by Galaxy Clusters}",
      journal = {\ssr},
         year = 2024,
        month = feb,
       volume = {220},
       number = {2},
          eid = {19},
        pages = {19},
          doi = {10.1007/s11214-024-01051-8},
archivePrefix = {arXiv},
       eprint = {2403.06245},
 primaryClass = {astro-ph.CO},
       adsurl = {https://ui.adsabs.harvard.edu/abs/2024SSRv..220...19N}
}

@ARTICLE{Wambs_rev_1998LRR.....1...12W,
       author = {{Wambsganss}, Joachim},
        title = "{Gravitational Lensing in Astronomy}",
      journal = {Living Reviews in Relativity},
         year = 1998,
        month = dec,
       volume = {1},
       number = {1},
          eid = {12},
        pages = {12},
          doi = {10.12942/lrr-1998-12},
archivePrefix = {arXiv},
       eprint = {astro-ph/9812021},
 primaryClass = {astro-ph},
       adsurl = {https://ui.adsabs.harvard.edu/abs/1998LRR.....1...12W}
}

@article{chwolson1924doppelsterne,
  author  = {Orest Chwolson},
  title   = {{\"U}ber eine m{\"o}gliche Form fiktiver Doppelsterne},
  journal = {Astronomische Nachrichten},
  volume  = {221},
  number  = {11},
  pages   = {329--330},
  year    = {1924},
  doi     = {10.1002/asna.19242211105}
}

@inbook{einstein1995zurich,
  author    = {Albert Einstein},
  title     = {The Collected Papers of Albert Einstein, Volume 4: The Swiss Years: Writings, 1912--1914},
  editor    = {Martin J. Klein and A. J. Kox and Jürgen Renn and Robert Schulman},
  publisher = {Princeton University Press},
  year      = {1995},
  chapter   = {Doc. 10 (``Einstein's Research Notes on a Generalized Theory of Relativity'')},
  pages     = {201--294},
  address   = {Princeton, NJ}
}

@article{Abbott2024_LensingO3,
  author = {{LIGO Scientific Collaboration} and {Virgo Collaboration} and {KAGRA Collaboration}},
  title = {Search for Gravitational-Lensing Signatures in the Full Third Observing Run of the LIGO--Virgo Network},
  journal = {Astrophysical Journal},
  volume = {970},
  pages = {191},
  year = {2024},
  eprint = {2304.08393},
  archivePrefix = {arXiv},
  primaryClass = {gr-qc}
}

@article{Janquart2023_LensingFollowup,
  author = {Janquart, Justin and Wright, M. and Goyal, S. and Chan, J. C. L. and Ganguly, A. and Garr{\'o}n, {\'A}. and Keitel, D. and Li, A. K. Y. and Liu, A. and Lo, R. K. L.},
  title = {Follow-up analyses to the O3 LIGO--Virgo--KAGRA lensing searches},
  journal = {Monthly Notices of the Royal Astronomical Society},
  volume = {526},
  number = {3},
  pages = {3832--3860},
  year = {2023},
  doi = {10.1093/mnras/stad2909}
}

@article{Janquart2021,
  author = {Janquart, J. and Hannuksela, O. A. and Haris, K. and Van Den Broeck, C.},
  title = {A fast and precise methodology to search for and analyse strongly lensed gravitational-wave events},
  journal = {Mon. Not. R. Astron. Soc.},
  volume = {506},
  number = {4},
  pages = {5430--5438},
  year = {2021},
  doi = {10.1093/mnras/stab1991}
}

@article{Janquart2023,
  author = {Janquart, J. and Haris, K. and Hannuksela, O. A. and Van Den Broeck, C.},
  title = {The return of GOLUM: improving distributed joint parameter estimation for strongly lensed gravitational waves},
  journal = {Mon. Not. R. Astron. Soc.},
  volume = {526},
  number = {2},
  pages = {3088--3098},
  year = {2023},
  doi = {10.1093/mnras/stad2838}
}

@article{Li2023,
  author = {Li, A. K. Y. and Lo, R. K. L. and Sachdev, S. and Chan, J. C. L. and Lin, E. T. and Li, T. G. F. and Weinstein, A. J.},
  title = {Targeted subthreshold search for strongly lensed gravitational-wave events},
  journal = {Phys. Rev. D},
  volume = {107},
  pages = {123014},
  year = {2023},
  doi = {10.1103/PhysRevD.107.123014}
}

@article{Lo2023,
  author = {Lo, R. K. L. and Maga\~{n}a Hernandez, I.},
  title = {Bayesian statistical framework for identifying strongly lensed gravitational-wave signals},
  journal = {Phys. Rev. D},
  volume = {107},
  pages = {123015},
  year = {2023},
  doi = {10.1103/PhysRevD.107.123015}
}

@article{Bianconi2023,
  author = {Bianconi, M. and Smith, G. P. and Nicholl, M. and Ryczanowski, D. and Richard, J. and Jauzac, M. and Massey, R. and Robertson, A. and Sharon, K. and Ridley, E.},
  title = {On the gravitational lensing interpretation of three gravitational wave detections in the mass gap by LIGO and Virgo},
  journal = {Mon. Not. R. Astron. Soc.},
  volume = {521},
  number = {3},
  pages = {3421--3430},
  year = {2023},
  doi = {10.1093/mnras/stad673}
}

@article{Chakraborty2026,
  author = {Chakraborty, A. and Mukherjee, S.},
  title = {Model-independent search discards faint lensed-pairs of gravitational wave events in the sub-threshold candidates of GWTC-4},
  journal = {arXiv e-prints},
  eprint = {2606.03346},
  year = {2026}
}

@article{Takahashi2003,
  author = {Takahashi, R. and Nakamura, T.},
  title = {Wave effects in gravitational lensing of gravitational waves from chirping binaries},
  journal = {Astrophys. J.},
  volume = {595},
  pages = {1039--1051},
  year = {2003},
  doi = {10.1086/377430}
}

@article{Dai2017_Population,
  author = {Dai, L. and Venumadhav, T. and Sigurdson, K.},
  title = {Effect of lensing magnification on the apparent distribution of black hole mergers},
  journal = {Phys. Rev. D},
  volume = {95},
  pages = {044011},
  year = {2017},
  doi = {10.1103/PhysRevD.95.044011}
}

@article{Wierda2021,
  author = {Wierda, A. R. A. C. and Wempe, E. and Hannuksela, O. A. and Koopmans, L. V. E. and Van Den Broeck, C.},
  title = {Beyond the detector horizon: Forecasting gravitational-wave strong lensing},
  journal = {Astrophys. J.},
  volume = {921},
  pages = {154},
  year = {2021},
  doi = {10.3847/1538-4357/ac1bb4}
}

@article{Meena2020,
  author = {Meena, A. K. and Bagla, J. S.},
  title = {Gravitational lensing of gravitational waves: wave nature and prospects for detection},
  journal = {Mon. Not. R. Astron. Soc.},
  volume = {492},
  pages = {1127--1134},
  year = {2020},
  doi = {10.1093/mnras/stz3509}
}

@article{ChenLu2026,
  author = {Chen, Z. and Lu, Y.},
  title = {Gravitational lensing of gravitational waves from astrophysical sources: theory, detection, and applications},
  journal = {arXiv e-prints},
  eprint = {2605.06321},
  year = {2026}
}

@article{Caliskan2024,
  author = {Caliskan, M. and Kumar, N. A. and Ji, L. and Ezquiaga, J. M. and Cotesta, R. and Berti, E. and Kamionkowski, M.},
  title = {Probing wave-optics effects and low-mass dark matter halos with lensing of gravitational waves from massive black holes},
  journal = {Phys. Rev. D},
  volume = {109},
  pages = {063032},
  year = {2024},
  doi = {10.1103/PhysRevD.109.063032}
}

@article{Oguri2018,
  author = {Oguri, M.},
  title = {Effect of gravitational lensing on the distribution of gravitational waves from distant binary black hole mergers},
  journal = {Mon. Not. R. Astron. Soc.},
  volume = {480},
  number = {3},
  pages = {3842--3855},
  year = {2018},
  doi = {10.1093/mnras/sty2145}
}

@software{alex_nitz_2022_6324278,
  author       = {Alex Nitz and
                  Ian Harry and
                  Duncan Brown and
                  Christopher M. Biwer and
                  Josh Willis and
                  Tito Dal Canton and
                  Collin Capano and
                  Thomas Dent and
                  Larne Pekowsky and
                  Andrew R. Williamson and
                  Soumi De and
                  Miriam Cabero and
                  Bernd Machenschalk and
                  Duncan Macleod and
                  Prayush Kumar and
                  Steven Reyes and
                  dfinstad and
                  Francesco Pannarale and
                  Sumit Kumar and
                  Thomas Massinger and
                  Márton Tápai and
                  Leo Singer and
                  Gareth S Cabourn Davies and
                  Sebastian Khan and
                  Stephen Fairhurst and
                  Alex Nielsen and
                  Shashwat Singh and
                  Koustav Chandra and
                  shasvath and
                  veronica-villa},
  title        = {gwastro/pycbc: v2.0.2 release of PyCBC},
  month        = mar,
  year         = 2022,
  publisher    = {Zenodo},
  version      = {v2.0.2},
  doi          = {10.5281/zenodo.6324278},
  url          = {https://doi.org/10.5281/zenodo.6324278}
}

@article{BullockBoylanKolchin2017,
  author = {Bullock, James S. and Boylan-Kolchin, Michael},
  title = {Small-Scale Challenges to the {$\Lambda$}CDM Paradigm},
  journal = {Annual Review of Astronomy and Astrophysics},
  year = {2017},
  volume = {55},
  pages = {343--387},
  doi = {10.1146/annurev-astro-091916-055313}
}

\end{document}